\documentclass[twocolumn]{aastex631}

\usepackage{array}
\usepackage{amsmath}
\usepackage{float}
\usepackage{blindtext}
\newcolumntype{?}{!{\vrule width 1.5pt}}
\usepackage{enumerate}
\usepackage{hyperref}
\usepackage{graphicx}
\expandafter\ifx\csname natexlab\endcsname\relax\fi
\providecommand{\url}[1]{\href{#1}{#1}}
\providecommand{\dodoi}[1]{doi:~\href{http://doi.org/#1}{\nolinkurl{#1}}}
\providecommand{\doeprint}[1]{\href{http://ascl.net/#1}{\nolinkurl{http://ascl.net/#1}}}
\providecommand{\doarXiv}[1]{\href{https://arxiv.org/abs/#1}{\nolinkurl{https://arxiv.org/abs/#1}}}

\shorttitle{ASASSN-14li}
\shortauthors{Ajay et al.}
\graphicspath{{./}{figures/}}
\newcommand\xmm{\textit{XMM-Newton}\xspace}

\begin{document}
\nolinenumbers
\title{Episodic X-ray Outflows from the Tidal Disruption Event ASASSN-14li}

    \author{Yukta Ajay}
    \affiliation{Department of Physics and Astronomy, Johns Hopkins University, 3400 N. Charles Street, Baltimore, MD 21218, USA}
    \affiliation{Indian Institute of Science Education and Research (IISER) Tirupati, Tirupati 517 507, Andhra Pradesh, India}	
    \author{Dheeraj R. Pasham}
    \affiliation{MIT Kavli Institute for Astrophysics and Space Research, Massachussetts Institute of Technology, Cambridge, MA 02139, USA}
    \author{Thomas Wevers}
    \affiliation{Space Telescope Science Institute, 3700 San Martin Drive, Baltimore, MD 21218, USA}
    \affiliation{European Southern Observatory, Alonso de C{\'o}rdova, 3107 Vitacura, 6730000, Santiago, Chile}
    \author{Eric R. Coughlin}
    \affiliation{Department of Physics, Syracuse University, Syracuse, NY 13244, USA}
    \author{Francesco Tombesi}
    \affiliation{Physics Department, Tor Vergata University of Rome, Via della Ricerca Scientifica 1, 00133 Rome, Italy}
    \affiliation{INAF – Astronomical Observatory of Rome, Via Frascati 33, 00040 Monte Porzio Catone, Italy}
    \affiliation{INFN - Rome Tor Vergata, Via della Ricerca Scientifica 1, 00133 Rome, Italy}
    \affiliation{Department of Astronomy, University of Maryland, College Park, MD 20742, USA}
    \affiliation{NASA Goddard Space Flight Center, Code 662, Greenbelt, MD 20771, USA}
    \author{Muryel Guolo}
    \affiliation{Department of Physics and Astronomy, Johns Hopkins University, 3400 N. Charles Street, Baltimore, MD 21218, USA}
    \author{James F. Steiner}
    \affiliation{Harvard-Smithsonian Center for Astrophysics, 60 Garden Street, Cambridge, MA 02138, USA}

    \correspondingauthor{Yukta Ajay}
    \email{yajay1@jh.edu}

\begin{abstract}
 \noindent ASASSN-14li is a low-redshift ({\it z=} 0.0206) tidal disruption event (TDE) that has been studied extensively across the entire electromagnetic spectrum, and has provided one of the most sensitive measurements of a TDE to-date. Its X-ray spectrum is soft and thermal (kT$\sim$ 0.05 keV) and shows a residual broad absorption-like feature between 0.6-0.8 keV, which can be associated with a blue-shifted O~VII (rest-frame energy 0.57 keV) resulting from an ultrafast outflow (UFO) at early times (within 40 days of optical discovery). By carefully accounting for pile-up and using precise XSTAR photo-ionization table models, we analyze the entire archival X-ray data from \xmm and track the evolution of this absorption feature for $\sim$4.5 years post disruption. Our main finding is that, contrary to the previous literature, the absorption feature is transient and intermittent. Assuming the same underlying physical basis (i.e. outflows) for the recurring absorption feature in ASASSN-14li, the outflow is seen to disappear and reappear multiple times during the first $\sim$2 years of its evolution. No observable spectral imprint is detected thereafter. While theoretical studies suggest the launch of outflows in the early phases of the outburst during the super-Eddington regime, the outflow's intermittent behavior for multiple years after disruption is unusual. We discuss this peculiar behavior within the context of varying  inner disk truncation, radiation pressure, and magnetically-driven outflow scenarios.
\end{abstract}

\keywords{black hole physics, tidal disruption, relativistic processes, X-rays, accretion}

\section{Introduction}
When a star approaches close enough to a massive black hole such that the tidal shear across its diameter exceeds its self-gravity, it will be disrupted in a stellar tidal disruption event \citep[TDE,][]{1988Natur.333..523R}. Such TDEs can produce panchromatic flares and are signposts of otherwise-dormant black holes. They provide clean setups to study the onset of accretion and any accompanying ejections from regions close to the black hole. Depending on the black hole mass \citep[e.g.,][]{2018MNRAS.478.3016W}, the initial phases of accretion onto the central black hole are predicted to follow near-Eddington or super-Eddington fallback rates \citep{1988Natur.333..523R, 1997ApJ...489..573L, 2009MNRAS.400.2070S, 2011MNRAS.415..168S}. Theoretical studies have predicted the launch of outflows during the super-Eddington accretion phase, which can manifest as observational signatures in the X-rays \citep{2009MNRAS.400.2070S, 2011MNRAS.415..168S, coughlin14}.

\par Located in the nearby galaxy PGC043234 (at a redshift of $z$=0.0206, and a luminosity distance of $d_{L}$= 90.3~Mpc), ASASSN-14li is an optically selected TDE discovered by the All-Sky Automated Survey for SuperNovae \citep[ASAS-SN, ][]{2014AAS...22323603S} on 22 November 2014 (MJD 56983.6) \citep{2014ATel.6777....1J}. Due to its close proximity, ASASSN-14li is among the most well-studied TDEs, and has been observed as part of several multiwavelength follow-up campaigns in the optical, UV, X-rays, radio, and the IR \citep{2015Natur.526..542M, 2016MNRAS.455.2918H, 2016ApJ...818L..32C, 2016ApJ...832L..10R, 2016ApJ...819L..25A, 2016Sci...351...62V, 2016ApJ...828L..14J, 2017ApJ...837L..30P}. The mass of the central black hole was found to be $\sim$10$^{6}$\(\textup{M}_\odot\), based on several independent lines of evidence \citep{2015Natur.526..542M,2017MNRAS.471.1694W, 2019Sci...363..531P, 2020MNRAS.492.5655M, 2023MNRAS.522.1155W}. 

\par
\begin{figure}[]
    \includegraphics[width=\columnwidth]{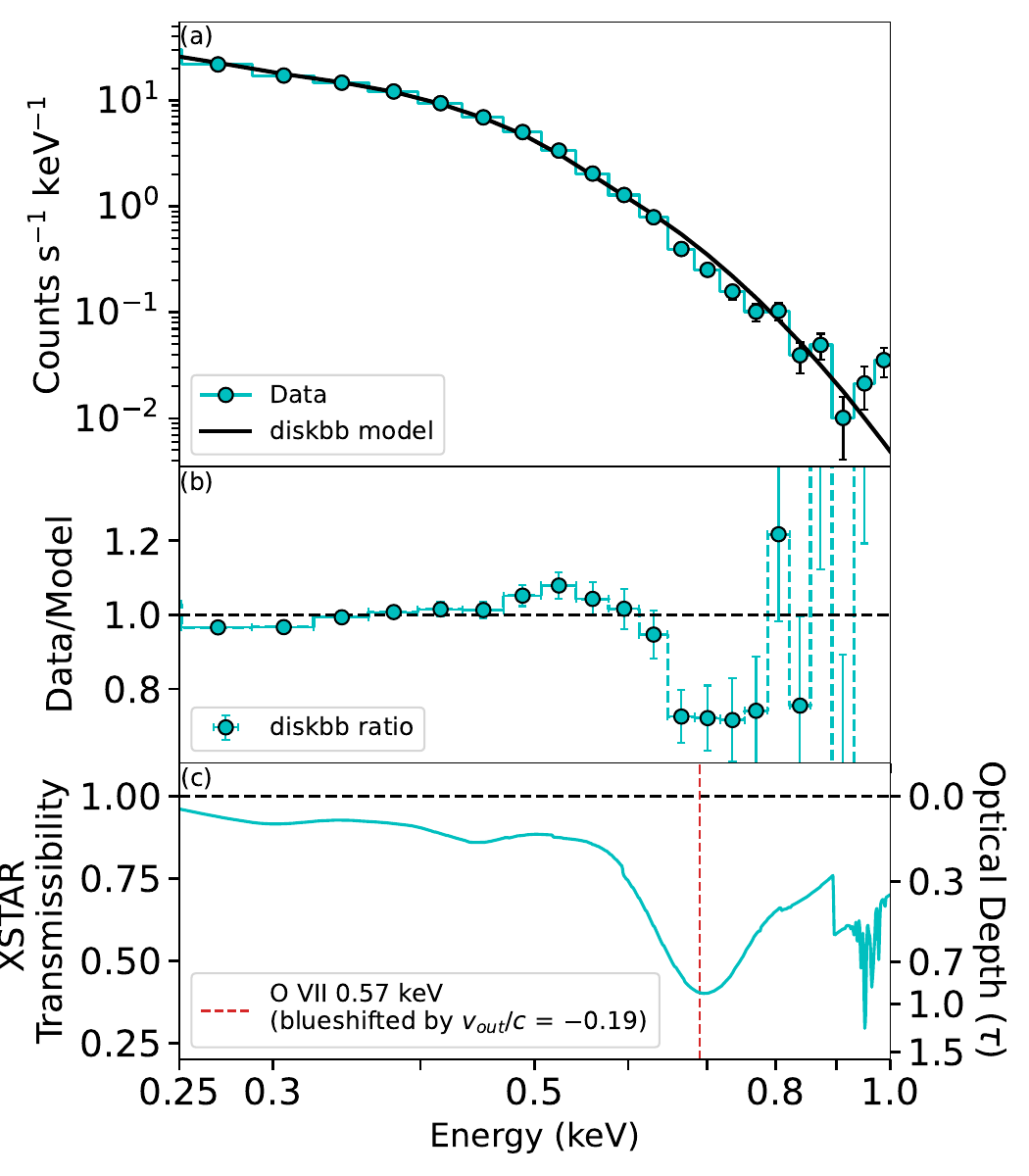}
       \caption{(a) X1 spectrum in the 0.25 - 1 keV bandpass and the best-fit disk blackbody model; (b) Ratio plot of the data to the best-fit disk blackbody model for X1 shows an absorption-like feature between 0.6 - 0.8 keV; (c) Best-fitted XSTAR absorption model for X1. The main feature can be associated with  O~VII line  (rest-frame 0.57 keV) blue-shifted by a outflow velocity of $\sim-0.19c$. The vertical axis to the left gives the transmissibility of the absorber/UFO (best-fitted XSTAR model), while the right axis shows its optical depth. The horizontal axis in all panels represents energies in the observer frame of reference.}
    \label{fig:abs_O7}
\end{figure}

 Early time observations taken within 40 days of the TDE discovery revealed a broad absorption feature at $\sim$0.7 keV in the soft X-ray spectrum, which was interpreted as a highly ionized ultrafast outflow \citep[UFO; e.g.,][]{2011ApJ...742...44T} with velocity $\sim$-0.2c \citep{2018MNRAS.474.3593K} (negative sign taken to represent blue-shifted velocities; also see Figure \ref{fig:abs_O7}c). Additionally, ASASSN-14li's X-ray power density spectrum (PDS) revealed a stable quasi-periodic oscillation (QPO) with a centroid frequency of $\approx$7.7 mHz \citep{2019Sci...363..531P}. This QPO was more coherent and had a root-mean-squared amplitude that was higher by at least a factor of 10 compared to QPOs seen in accreting stellar-mass black holes, and was attributed to activity close to the event horizon of the (rapidly spinning) black hole \citep{2019Sci...363..531P}.

\par Transient radio synchrotron emission was also detected from ASASSN-14li. Several competing models have been proposed to explain the  nature of this radio emission \citep{2016ApJ...819L..25A, 2016ApJ...829...19V, 2016ApJ...827..127K, 2018ApJ...856....1P}, of which the adiabatically-expanding jet model \citep{2018ApJ...856....1P} identifies the coupling of accretion rate and the jet power to be approximately linear, and attributes the delayed variability in the radio (following variability in the soft X-rays) to an internal emission mechanism within the freely-expanding jet. Further, the delayed IR variability (with respect to the optical/UV peak) in ASASSN-14li confirmed the possibility of TDE dust echoes, which result from the absorption and re-emission of optical and UV radiation by dust grains in the vicinity of the black hole \citep{2016ApJ...828L..14J}.

\par Here we analyze ASASSN-14li's time-resolved X-ray spectra from archival {\it XMM-Newton} data to track the broad absorption feature detected in the soft X-rays over a temporal baseline of 4.5 years. Specific details of data reduction procedures and model parameters are provided in the Appendix. All uncertainties presented in this work represent a 90\% confidence range, unless stated otherwise. 
\begin{table*}
\centering
\begin{tabular}{|c||c|c|p{1.5cm}|p{1.5cm}|p{1.5cm}|p{2.2cm}|p{1.5cm}|c|}
\hline
Observation  & ObsID & Date & Start time (MJD) & End time (MJD) & Net exposure (ks) & Combined GTI exposure (ks) & Effective count rate (cps) & Absorber?\\ [0.5ex]
 \hline\hline
  X1 & 0694651201 & 2014-12-06 & 56997.98 & 56998.23 & \centering{21.2} & \centering{7.7} & \centering{3.503} & {\bf Yes}\\ [1ex]
 \hline
 X2 & 0722480201 & 2014-12-08 & 56999.54 & 57000.62 & \centering{93.9} & \centering{10.9} & \centering{1.761} & {\bf Yes}\\ [1ex]
 \hline
 X3 & 0694651401 & 2015-01-01 & 57023.89 & 57024.17 & \centering{23.7} & \centering{16.0} & \centering{1.703} & {\bf Yes}\\ [1ex]
 \hline
 X4 & 0694651501 & 2015-07-10 & 57213.25 & 57213.51 & \centering{22.4} & \centering{15.0} & \centering{0.832} & No\\[1ex]
 \hline 
 X5 & 0770980101 & 2015-12-10 & 57366.49 & 57367.59 & \centering{95.4} & \centering{35.7} & \centering{0.652} & {\bf Yes}\\ [1ex]
 \hline
 X6 & 0770980501 & 2016-01-12 & 57399.18 & 57399.28 & \centering{8.9} & \centering{3.1} & \centering{1.803} & No\\   [1ex]
 \hline 
 X7 & 0770980601 & 2016-06-04 & 57543.98 & 57544.13 & \centering{13.4} & \centering{9.1} & \centering{0.676} & {\bf Yes}\\   [1ex]
 \hline
 X8 & 0770980701 & 2016-12-04 & 57726.60 & 57726.71 & \centering{9.7} & \centering{6.3} & \centering{0.354} & {\bf Yes}\\   [1ex]
 \hline
 X9 & 0770980801 & 	2017-06-08 & 57912.05 & 57912.18 & \centering{11.9} & \centering{8.1} & \centering{0.196} & No\\   [1ex]
 \hline
 X10 & 0770980901 & 2017-12-05 & 58092.37 & 58092.65 & \centering{24.7} & \centering{2.5} & \centering{0.128} & No\\   [1ex]
 \hline 
 X11 & 0770981001 & 2018-07-05 & 58304.69 & 58304.82 & \centering{11.9} & \centering{7.7} & \centering{0.071} & No\\   [1ex]
 \hline
 X12 & 0810200501 & 2018-12-10 & 58462.57 & 58462.89 & \centering{29.9} & \centering{6.5} & \centering{0.054} & No\\   [1ex]
 \hline
 X13 & 0810200701 & 2019-06-06 & 58640.82 & 58641.16 & \centering{31.4} & \centering{25.3} & \centering{0.051} & No\\  [1ex]
 \hline
\end{tabular}
\caption{Summary of \xmm observations including the final exposure times after accounting for background flaring and combining the GTIs, and the 0.25-1 keV background-subtracted count rates for X1-13.}
\label{table1}
\end{table*}
\\
\\
\section{Observations and Data reduction}
\label{section:data}
We have used all the thirteen archival, publicly available observations (hereafter referred to as X1, X2, X3, ..., X13) taken by {\it XMM-Newton}'s EPIC-pn instrument between December 2014 and June 2019. Previous works using data captured by {\it XMM-Newton}'s RGS, EPIC-pn and MOS cameras confirmed that the absorption feature in ASASSN-14li's X-ray spectrum was not an outcome of pile-up, but was indeed due to the presence of a highly ionized UFO at those times \citep{2018MNRAS.474.3593K}. In this work, we use data from the EPIC-pn instrument alone due to its larger effective area and higher sensitivity compared to the MOS detectors. We implemented the standard procedures for data reduction and analysis using the {\it XMM-Newton} Science Analysis System (SAS v.20.0.0) and the most recent calibration files. Since we were concerned with the EPIC-pn data, we executed {\tt epproc} to process the relevant observation data files (ODFs) and obtained  cleaned event lists.

\par Prior to extracting the X-ray spectra, we accounted for effects of photon pile-up \citep{2015A&A...581A.104J} in observations X1-5 by considering an annular source region of a fixed outer radius (${33"}$ = 660 physical units) centered on the source coordinates (J2000 RA=
12:48:15.23, DEC= +17:46:26.22). The inner exclusion radii were selected after quantifying the pile-up in each case by following the procedure outlined on
\href{https://www.cosmos.esa.int/web/xmm-newton/sas-thread-epatplot}{{\it XMM-Newton}'s data analysis pages}. The source was faint during observations X6-13 and had negligible pile-up, so we refrained from excluding any area in the selected source region to prevent loss of data. A description of the procedure we followed to mitigate pile-up is elaborated in Appendix \ref{section:pileup}. Background spectra were extracted using events from two circular regions with a radii of ${45"}$ near the source. 

\par Furthermore, we define good time intervals (GTIs) as the intervals of time with negligible background flaring, i.e., total field of view count rates typically  $<=$0.4 cps in the 10-12 keV energy range (PI$>$10000 \&\& PI$<$12000). A summary of the final exposure times for each observation after accounting for flaring is given in Table \ref{table1}.

\par We generated the redistribution matrix files (RMFs) and the ancillary response files (ARFs) using the {\tt rmfgen} and {\tt arfgen} tools available in XMM-SAS. For X-ray spectral analysis, we opted to use only the single-event spectra (filtered by setting PATTERN==0). We binned the spectra to include a minimum of 1 count per bin using an optimal binning scheme described in \cite{2016A&A...587A.151K}, and used the C-statistic to evaluate the model fits. Subsequent X-ray spectral analysis was performed using {\tt XSPEC} v12.12.1 \citep{1996ASPC..101...17A}. The background begins to dominate beyond 1 keV for most observations (see Appendix Figure \ref{fig:srcbkg}), and we therefore chose to fit the spectra over the 0.25-1 keV bandpass.

\section{X-ray spectral analysis}
\label{section:models}
\subsection{Modeling with thermal components}
We began by fitting all the thirteen observations using a disk blackbody since ASASSN-14li's spectrum is soft and thermal \citep[][]{2015Natur.526..542M, 2016MNRAS.455.2918H, 2017MNRAS.466.1275B, 2018MNRAS.474.3593K}. We used the {\tt TBabs*zTBabs*zashift*clumin*diskbb} model in {\tt XSPEC}, taking {\tt TBabs} to be Milky Way's neutral column density along our line-of-sight, fixed at $2\times10^{20}$ cm$^{-2}$ \citep[][]{2016A&A...594A.116H}. The host galaxy's neutral column was allowed to be a free parameter in X1-8, and fixed (at {\tt zTBabs} $=0$ cm$^{-2}$) in X9-13, where the best-fit column densities were consistent with zero. {\tt diskbb} is a thermal model  \citep[e.g., ][]{1986ApJ...308..635M} that has two free parameters, the temperature near the inner accretion disk and the normalization (see \href{https://heasarc.gsfc.nasa.gov/xanadu/xspec/manual/node163.html}{\it XSPEC documentation}\footnote{\href{https://heasarc.gsfc.nasa.gov/xanadu/xspec/manual/node163.html}{https://heasarc.gsfc.nasa.gov/xanadu/xspec/manual/node163.html}} for details).

\begin{figure*} 
    \centering
    \includegraphics[scale=0.6]{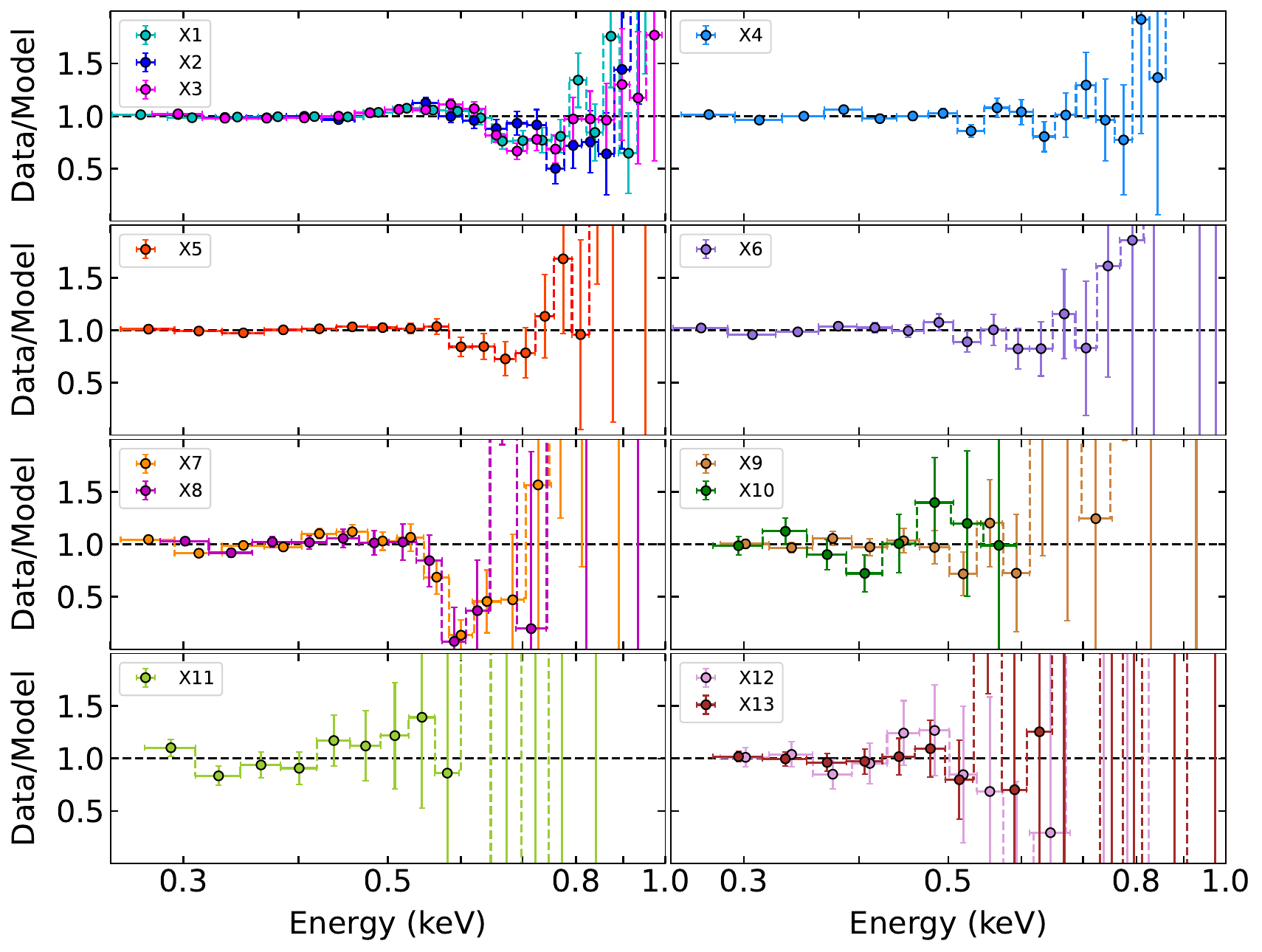}
    \caption{Data/model ratios for a disk blackbody model in the 0.25-1 keV energy range (observer frame) shows an absorption-like feature between 0.6-0.8 keV in X1-3, X5, X7-8, indicative of the presence of ionized outflows in these observations. Another version of this same data with an extended y-scale can be found in supplementary digital file ``{\small \tt diskbb$\_$ratios$\_$extended$\_$yscale.pdf}".}
    \label{fig:fig1}
\end{figure*}

\par In Figure \ref{fig:abs_O7}, we show X1's spectrum and the best-fit disk blackbody model for this observation. The resulting ratio of the data and the best-fit disk blackbody model shows a broad absorption-like feature between 0.6-0.8 keV, similar to the spectral feature discussed in \cite{2018MNRAS.474.3593K} as evidence for an outflow. Photo-ionization modeling with an appropriate XSTAR model (discussed later) and correcting for the source redshift indicates that the absorption feature is an outcome of O~VII (0.57 KeV) that is blue-shifted by an ionized UFO of velocity $\sim-0.2c$. In Figure \ref{fig:fig1}, we show the ratios of the data to the best-fit disk blackbody model for all observations, from which we see that X2, X3, X5, X7, and X8 also show a similar absorption line-like feature. With the above model, the ratio of C-stat/degrees of freedom (d.o.f) for these observations is $>$2, implying that a single disk blackbody cannot fully describe the data in the above cases. The C-stat/d.o.f ratio is better in the remaining observations, with a maximum value of $\approx$1.8 in X4, and ratios closer to 1 in X6 and X9-13 (see Appendix Table \ref{table:diskbb}). Observations X9-13 were obtained at late times when the source  faded appreciably as the count rates dropped by a factor of $\approx$18 in X9 compared to X1, with even lower count rates in the successive observations X10-13. A disk-blackbody model adequately describes these late-time data. Despite the comparatively poorer statistical quality of these spectra, we rule out the presence of an absorption feature in X9-13 (discussed in Section \ref{subsection:X9to13NoDet}). 

\begin{figure*} 
    \centering
    \includegraphics[scale=0.6]{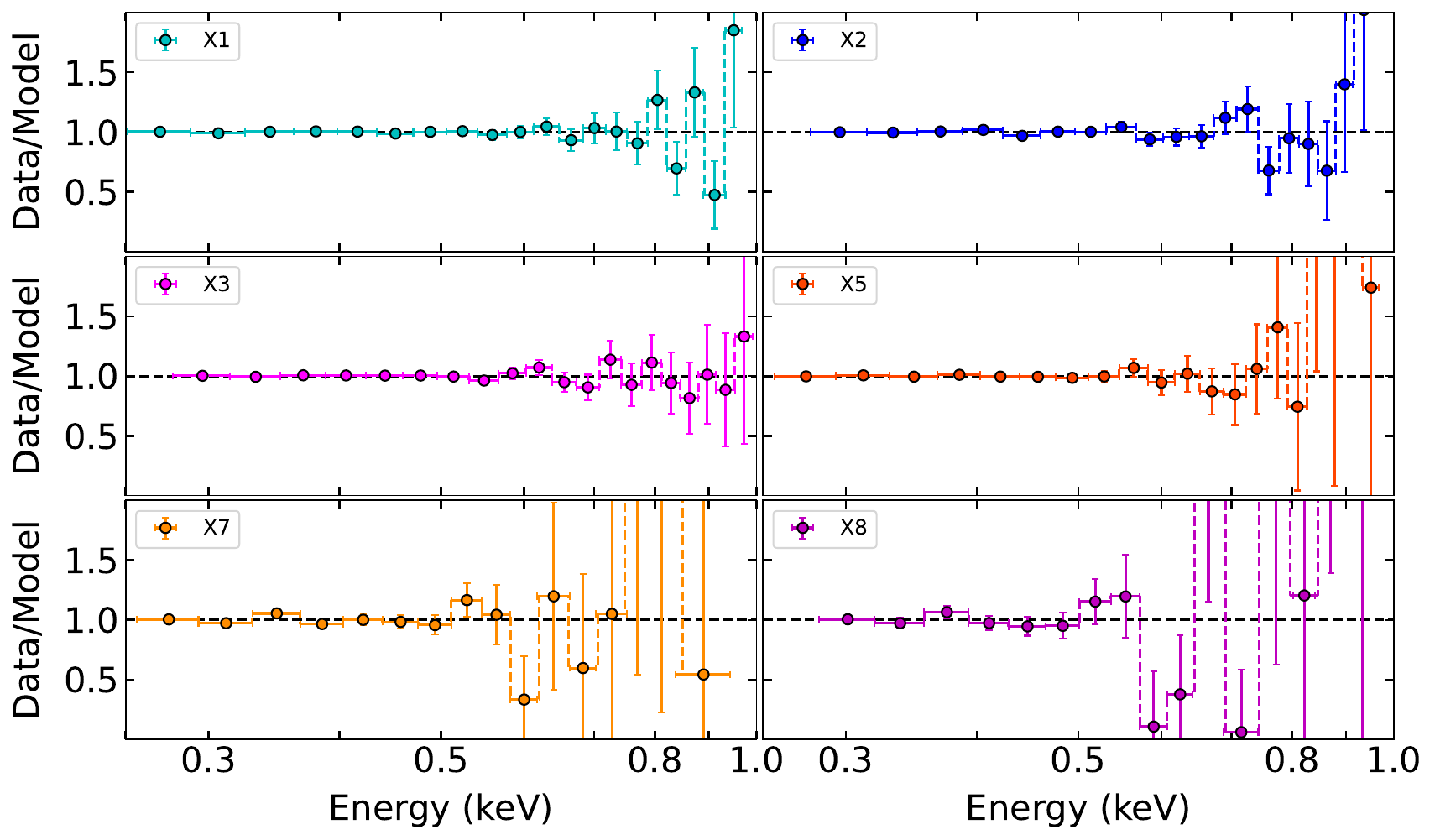}
    \caption{Data/model ratios in the 0.25-1 keV energy range (observer frame) for a model with disk blackbody and an ionized absorber (XSTAR) component for X1-3, X5, X7+8, where the broad absorption feature is now accounted for.}
    \label{fig:fig2}
\end{figure*}

\par Because \cite{2018MNRAS.474.3593K} used a dual disk blackbody model, we next attempted to fit the data in X1-8 with two disk blackbody components to check whether including a second disk blackbody at a different temperature improved the spectral fits. \cite{2018MNRAS.474.3593K} reported the best-fit temperature of one disk blackbody to be  between $\sim 0.04 - 0.06$ keV, and the second disk blackbody to be at a higher temperature (typically $ > 0.15$ keV) for X1-6. We followed a similar approach and fit X1-8 spectra using the {\tt TBabs*zTBabs*zashift*(diskbb+diskbb)} model. The best-fit temperature of one of the disks was equal to the temperature from the single disk blackbody model within errors, while the second disk temperature was consistent with $\approx$0K or with the first disk blackbody temperature, within errors. The inclusion of a second disk blackbody did not significantly improve the C-statistic/d.o.f in comparison to the respective ratios from a model with only one disk blackbody component. This suggests that the second disk blackbody at a higher temperature is redundant. We discuss the reasons for this discrepancy further in Section \ref{subsection:comparison} and show that the second disk blackbody component is in fact fitting the hard excess at higher energies that is a manifestation of pileup in the case of \cite{2018MNRAS.474.3593K}.

\subsection{Modeling with thermal component + photo-ionized absorber}
\label{subsection:diskufo}
\par With the above spectral analysis, it is evident that we need an additional component in our models to account for the absorption-like feature in ASASSN-14li's spectra. To constrain the parameters of the absorber, we generated photo-ionization models using XSTAR code v2.39 \citep{2001ApJS..133..221K}. We considered a constant density shell of $10^{10}$~cm$^{-3}$, although the precise value of the density is not strictly important for a geometrically-thin shell because the code would simply scale the distance in order to obtain the same ionization parameter. All elemental abundances were fixed to solar values. A velocity broadening of 30,000 km~s$^{-1}$ was chosen to provide the best-fit to the observed broad absorption feature, even though the energy resolution of the data below 1 keV is rather limited. Such a high velocity broadening is most likely indicating a rotation of the absorbing material close to the black hole or a large line-of-sight outflow velocity shearing. Given the lack of additional information (such as disk inclination angle and wind geometry), we do not further investigate the physical origin of this parameter but consider it only for a phenomenological fit.

\par The XSTAR tables employed in our analysis are specific to the approximate disk temperature and luminosity of each observation for better precision, and categorized into three groups for high (Table-H, for $kT$ (temperature)=0.06 keV, log($L$)=44.5 ergs s$^{-1}$ for X1-3); medium (Table-M, for $kT$=0.05 keV, log($L$)=44.2 ergs s$^{-1}$ for X4-6); and low (Table-L, for $kT$=0.055 keV, log($L$)=44.0 ergs s$^{-1}$ for X7-8) temperature and luminosity values. Here, $L$ represents the integrated 13.6 eV to 13.6 keV luminosity. These XSTAR tables can be accessed as supplementary digital files\footnote{Supplementary digital files: {\small \tt mtable$\_$14li$\_$H.fits} (Table-H); {\small \tt mtable$\_$14li$\_$M.fits} (Table-M); and {\small \tt mtable$\_$14li$\_$L.fits} (Table-L).}. The tables yield three properties of the ionized absorber: (i) the equivalent (ionized) hydrogen column density along our line of sight, $N_H$ (sampled in the wide range of $10^{19}$-$10^{23}$ cm$^{-2}$); (ii) the ionization parameter in logarithmic scale, log($\xi$) (in the range 2 to 4 for Table-H, and 2 to 6 for Table-M and L); and (iii) the line of sight velocity, $v_{out}$ (in units of $c$, negative values indicate blue-shifted velocities). 

\par We fit the spectra for X1-8 with the {\tt TBabs*zTBabs*zashift*XSTAR*clumin(diskbb)} model in {\tt XSPEC}, and focused on the best-fit values for the disk temperature, luminosity, and XSTAR parameters. We present the data/model ratios for cases with an absorption feature in Figure \ref{fig:fig2}, and find the following trends by modeling the spectra with a disk blackbody and the respective XSTAR table components:

\begin{enumerate}[(i)]
    \item X1, X2 and X3 are well described with the XSTAR Table-H and a disk blackbody, resulting in approximately flat data/model ratios and reduced C-stat values close to unity in all three cases. The temperature and luminosity values match with the input XSTAR table values (within errors), which validates the accuracy of the XSTAR photoionization tables used in our analysis and confirms that these tables can properly account for the absorption signature seen in the early-time observations. 
    
    \item Among X4, X5, and X6, we have already seen that X5 shows a broad feature in its X-ray spectrum, and that X4 and X6 can be described by a simple disk blackbody model. In X5, the inclusion of the XSTAR Table-M along with a disk blackbody component (20/14) resulted in a notable improvement in the C-stat/d.o.f ratio compared to a model containing only a disk blackbody (33/17), However, using the Table-M for X4 and X6 does not improve the data/model ratios, and indicates that  ionized absorbers were likely absent or highly diminished in both these instances, or alternatively, out of our line-of-sight. Although we find that the absorption feature is absent in X6, it is important to note that the number of counts in X5 are higher than in X6 by a factor of $\approx$4. As a quick check, we fixed the log($\xi$) value to its lower hard-limit (= 2) and the outflow velocity to be $0c$ for X4 and X6 to obtain  upper limits on their ionized column densities. On performing the spectral fitting, the temperature and luminosity values were close to the Table-M values in both cases, which gave us confidence to still apply this model. The ionized column values for X4 and X6 have been represented as upper limits in Figure \ref{fig:ufo_timeevol}, and are lower than in the cases when the spectral feature was present. 
    
    \item For X7 and X8, we fit the spectra with a disk blackbody and XSTAR Table-L by fixing the host galaxy neutral hydrogen column (at {\tt zTBabs} $=0$ cm$^{-2}$) as it was consistent with zero. Agreement with the XSTAR Table-L suggests the reappearance of the absorption feature in the X7 and X8 spectra between $\sim$1.5-2 years after ASASSN-14li's optical discovery. The derived temperatures for X7 and X8 were slightly higher in comparison to the temperatures from the disk blackbody model (where the temperatures were $\approx$0.045 keV). This was likely caused by a slight change in the shape of the thermal continuum with the inclusion of the XSTAR component, and we therefore used a Table-L designed for a temperature $\approx$0.055 keV to account for this temperature increase in X7 and X8.  
\end{enumerate}

\subsection{Statistical tests and model selection}
\label{subsection:tests}
To verify that the transient absorption features we report in ASASSN-14li are not due to noise, we used two independent statistical tests: Akaike Information Criterion \cite[AIC;][]{aic} and the {\tt simftest} XSPEC tool. $\Delta$AIC can be used to test whether a more “complex” model (here, with XSTAR) is favored by quantifying the improvement due to the inclusion of the additional component. We applied this test to X1-8. In all observations where we find the feature to be present, the more complex model (with XSTAR+diskbb) is favored over the simpler model (with only diskbb), thus indicating the presence of absorption features. A more detailed description of this test can be found in Appendix \hyperref[section:stats]{B}. 

We further used the {\tt simftest} tool in XSPEC to estimate the probability that the absorption feature in the spectra results from statistical noise in X1-8. Results from this test have been shown as a continuation of Appendix \hyperref[section:stats]{B}. This probability is non-negligible only in X4 and X6, i.e. the same observations where we show that the spectral feature is absent, meaning that the remaining observations in X1-8 indeed have a statistically significant absorption feature ($>99\%$ significance) in their spectra. 

\par  With the above tests, we conclude that the absorption features are present in X1-3, X5, X7, and X8, and take the best fit model for these observations to be one with both {\tt diskbb} and XSTAR. We consider a model with {\tt diskbb} alone as best-fit for the remaining observations. Given that this spectral signature has been associated with (ultrafast) outflows in the early epochs (X1-3), we can assume the same physical basis for this feature in the subsequent epochs and attribute it to outflows at late times. We operate under this assumption of transient, late-time outflows to get limits on the blue-shifted outflow velocity, and to constrain the ionized column density and ionization parameter in X5, X7, and X8.

\subsection{No evidence for an absorption feature in X9-13}
\label{subsection:X9to13NoDet}
Given the limited statistics in X9-13, it became increasingly difficult to identify clear spectral features in these late time spectra. So, we predicted the expected number of counts in X9-13 if we were to observe a statistically significant absorption feature. Using the best-fit continuum parameters for X9 and the best-fit XSTAR parameters from X8, we simulated a spectrum using the {\tt fakeit} tool in {\tt XSPEC}. This simulated spectrum, which mimics the presence of an absorption feature in X9, had 1/3 the observed counts. In other words, the count rate would have been lower by a factor of 3 if the feature was present in X9. Similarly, the expected photon counts was less than half the observed counts in X10-12, again pointing towards the absence of the spectral feature of similar strength as X8. In X13, although the predicted number of counts was roughly 80\% of the observed counts, the data quality was very poor to successfully deduce the presence of an absorption feature. In all, the predicted number of counts in X9-13 with an absorber along our line-of-sight are  lower than the observed counts, suggesting the absence of the spectral signature, and thereby an ionized absorber (like outflows) as strong as the one in X8 in these observations. 

\subsection{No powerlaw or corona emission even after 2 years following optical discovery}
Powerlaw emission generally signifies non-thermal X-ray coronal emission, and is sometimes seen in TDEs after a few months or years following optical discovery \citep[e.g.,][]{2021ApJ...912..151W, 2022ApJ...937....8Y, 2022ATel15724....1A}. In ASASSN-14li, best-fit powerlaw indices in the late time observations X9-13 tended to unphysical values ($\Gamma > 8$ in most cases) when these spectra were individually fit with either {\tt (diskbb + powerlaw)} or {\tt powerlaw} alone, indicative of no underlying coronal emission. To verify this, X9-13 were jointly fit with {\tt TBabs*zTBabs*zashift(diskbb+ clumin*powerlaw)} to constrain the powerlaw luminosity. We froze the photon index at $\Gamma=2$, the approximate expected value from coronal emission, while the {\tt diskbb} parameters were independent and variable in each case. We assessed the powerlaw strength by comparing the resulting powerlaw luminosity (from {\tt diskbb + clumin*powerlaw} model) with the thermal disk luminosity (no powerlaw, {\tt clumin*diskbb}). The fractional strength of the coronal emission in comparison to the blackbody emission was $< 3\%$ of the total luminosity in the 0.3 to 10 keV range, and $< 0.03\%$ between 0.0136 to 13.6 keV energies. This points towards two possible scenarios - either that an X-ray corona had not formed even 2 years following discovery, or alternatively, that the corona might have been soft and thermal, indicative of a weak /subdominant state of the corona, akin to when X-ray binaries undergo outbursts \citep[][]{2006ARA&A..44...49R}.

\subsection{The X-ray feature cannot be explained with a slow-moving warm absorber}
In principle, the absorption feature seen in the ASASSN-14li spectra could arise from warm absorbers moving into our line-of-sight \citep[e.g.,][]{2000A&A...354L..83K, 2013HEAD...1310106T}. We test this scenario by fitting a warm absorber (WA) XSTAR table appropriate for an ionized absorption layer with a low velocity dispersion of $\sim$100 km~s$^{-1}$ far away from the source. Using the {\tt TBabs*zTBabs*zashift*WA*(diskbb)} model, we found that at all times when the broad absorption feature is detected, it persists even after the inclusion of the WA component. This suggests that the early- and late-time signatures in ASASSN-14li's X-ray spectra cannot be explained with distant, slow moving warm absorbers. 

\subsection{Comparison with previous works}
\label{subsection:comparison}
\cite{2018MNRAS.474.3593K} studied ASASSN-14li's X-ray data from X1-6 and reported the presence of a UFO in X1-3 which they claimed disappears or diminishes in X4-6. While we too find the UFO signature to be present in X1-3, our work suggests that the outflow, with considerably lower velocities at later times is transient during X4-6 (see Figure \ref{fig:fig1}). This difference can primarily be attributed to a careful consideration of pile-up effects (in all X1-5). We have also used a slightly wider energy band-pass of 0.25-1 keV at the lower end of the energy range \citep[0.3-1 keV in][]{2018MNRAS.474.3593K}, given that ASASSN-14li is a soft X-ray source, and maintained 1 keV as the upper energy limit, which allowed us to track the extended feature beyond 0.8 keV. Furthermore, we have binned the spectra using the optimal binning scheme of \cite{2016A&A...587A.151K}.

\par Next, \cite{2018MNRAS.474.3593K} used a model comprised of an absorbing outflow component and two disk blackbodies for X1-3, and only two disk blackbodies (no outflow) for X4-6, but had not adequately accounted for pile-up effects. While they argue that single pixel events are not piled-up or at least does not affect the PN spectra in X1-6, pattern distributions for X1-13 shown in the Appendix Figures \ref{fig:epat_w_pileup_corr}-\ref{fig:epat_no_pileup_corr} demonstrate that pile-up is indeed prominent even in single pixel events, which we correct for by excising central circular regions from X1-5 datasets. In our prior discussion on the two disk blackbody model, we showed that there is no need of the higher temperature disk blackbody in these data. This suggests that accounting for pile up effects corrected for the shift towards higher energies in the spectral energy distribution, meaning that the higher temperature disk blackbody was likely modeling the hard excess at higher energies caused by photon pile-up. While pile-up did not affect the X6 spectrum in our analysis, it was possible to model this spectrum with a single disk blackbody (C-stat/d.o.f = 15.3/16).

\par  To identify the transition line associated with the absorption feature in the spectrum, one can use the relation between the rest-frame transition energy ($E_{rf}$) and the observed transition energy ($E_{obs}$) given by $E_{rf} = E_{obs}\times (1+z)\times (1+z_{out})$, where $z$ is the source redshift, and $z_{out}$ is the outflow redshift ($\sim -0.19$ for X1). \citet{2018MNRAS.474.3593K} had reported the observed absorption feature to be a blue-shifted O~VIII line ($E_{rf}=0.653$ keV). Using the above equation, an O~VIII feature would have been observed at $E_{obs}\sim0.78$ keV, which is not the case (see Figure \ref{fig:abs_O7}). Instead, we find that the feature in the observed energy range $E_{obs} \approx 0.67-0.7$ maps to a rest-frame energy range $E_{rf} \approx 0.56-0.585$ and therefore infer that the strongest transition corresponds to an O~VII line at $E_{rf}$=0.57 keV per the results of the XSTAR absorption model. As a sanity check, we plot the observed line for a rest-frame transition energy $E_{rf}$ = 0.57 keV as the red line in Figure \ref{fig:abs_O7}c, and we see that this line aligns with the observed feature.

\par Finally, \citet{2018MNRAS.474.3593K} fitted a P-Cygni profile which requires both emission and absorption to model the observed feature. As we have
demonstrated, the emission component is not necessary, and a statistically-acceptable fit can be achieved (see Figure \ref{fig:fig2}) using only 
an absorption component. The apparent broad emission line-like hump around 0.55 keV is because the absorber distorts the 
continuum.

\section{Discussion}
\label{section:discussion}
By analyzing ASASSN-14li's X-ray spectra taken over 4.5 years following its optical discovery, we show the presence of a statistically significant absorption feature at late times up to 2 years (in X7 and X8) after its optical detection, which we associate with late-time outflows. The outflow is transient and intermittent (in X4-6), disappearing in X4 (6 months following optical discovery), after which it reappears in X5 (1 year following optical discovery), and disappears in X6 (1 month after X5). 

\begin{figure}
    \includegraphics[width=\columnwidth]{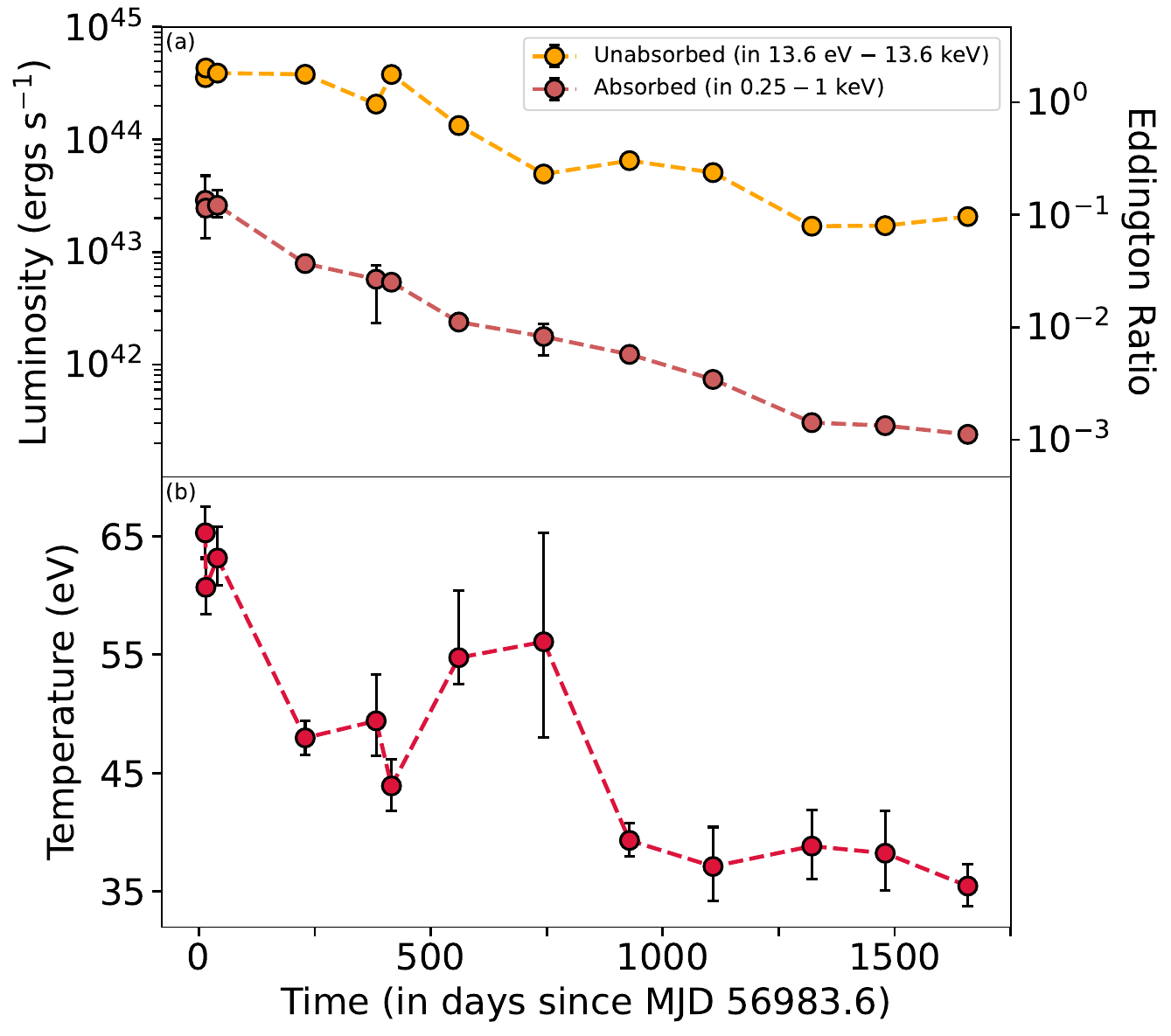}
    \caption{\small{Time evolution of temperature and luminosity using the respective best-fit models for X1-13, viz, disk blackbody model for X4, X6, and X9-13, and disk blackbody + UFO (XSTAR) for X1-3, X5, X7-8. Errors bars along x-axis represent the duration over which parameter values were calculated; (a) Model predicted unabsorbed bolometric luminosity (0.136 - 13.6 keV) and the observed luminosity (0.25 - 1 keV) are seen to decrease over time. The y-axis to the right gives the Eddington ratio calculated for ASASSN-14li's black hole mass $\sim10^{6.23}$\(\textup{M}_\odot\) \citep{2017MNRAS.471.1694W} for the corresponding luminosities on the vertical left-axis. The Eddington ratio in the initial observations was $>1$, indicating a super-Eddington accretion phase; (b) Best-fit disk temperatures decrease over time, with an exception in X7-8, where the slight rise in temperature can be attributed to a change in the continuum shape due to the presence of an absorbing outflow in these observations.}}
    \label{fig:tempevol}
\end{figure}

\subsection{Exploring mechanisms influencing outflow parameters}
\par
\begin{figure}
    \includegraphics[width=\columnwidth]{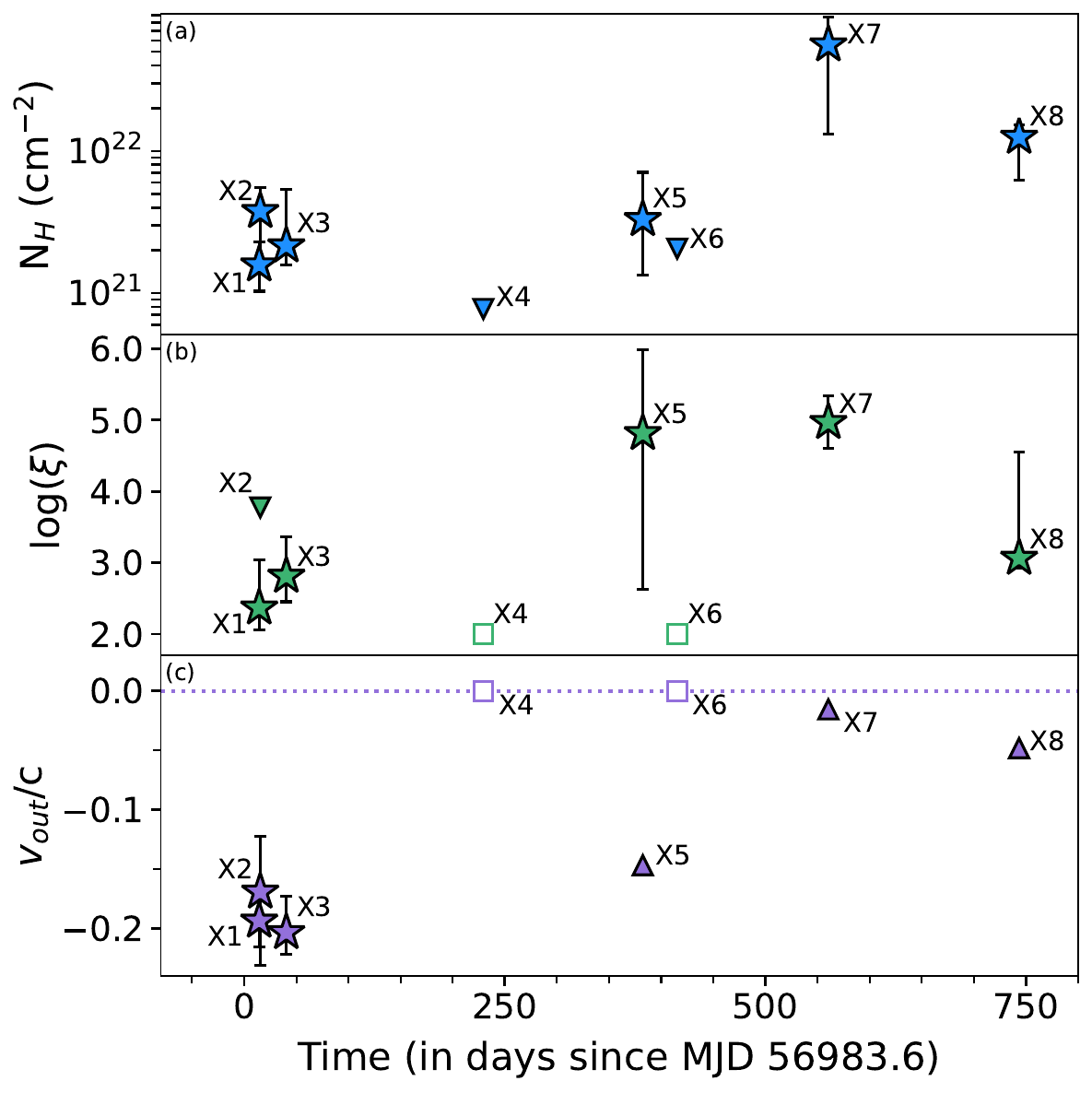}
    \caption{\small{Time evolution of outflow parameters for X1-8 using the disk blackbody + UFO (XSTAR) model -- (a) time evolution of the ionized column density ($N_{H}$ in cm$^{-2}$) for X1-8 in logarithmic scale; (b) time dependence of the ionization parameter ($\xi$, in logarithmic scale) for X1-8; and (c) time evolution of the blue-shifted outflow velocity for each observation X1-8 in units of $c$ ($v_{out}/c$). Star markers represent parameter values with errors, triangle markers with downward and upward arrows represent upper and lower limits for a particular parameter value respectively, and open square markers for X4 and X6 represent fixed parameter values for log($\xi$)=2 and $v_{out}=0c$ to obtain upper limits on the ionized columns in the respective cases.}}
    \label{fig:ufo_timeevol}
\end{figure}

 The time-evolution of outflow parameters in X1-8 is shown in Figure \ref{fig:ufo_timeevol}. The outflow velocity decreases with time:  $\sim -0.2$c in X1-3, $-0.15$c (upper limit) in X5, and roughly between -0.02c to -0.05c (upper limits) in X7-8.  Further, the decrease in outflow velocity is accompanied by an increase in the line-of-sight ionized column density by an order in magnitude in the later observations (particularly X7 and X8). This decrease in outflow velocity could indicate the truncation of the inner disk radius, or that successive outflows were launched from farther away in the disk. To investigate the latter scenario, we can compare the outflow speed to the escape speed at a given distance from the black hole. 
Specifically, if the gas is accelerated by the wind-driving mechanism on short timescales relative to the freefall time\footnote{If the mechanism responsible for driving the gas outward, which for ASASSN-14li is likely related to the interaction between the gas and the radiation field generated through accretion (see the discussion later in the present section), acted on timescales substantially longer than the freefall time, the gas would respond adiabatically/quasi-hydrostatically. In this case, the majority of the energy associated with the wind-driving mechanism would be lost to establishing a new equilibrium, rather than unbinding material.}, this distance likely correlates with the minimum radius from which the outflow is launched, as we require that the outflow speed $v_{out}$ satisfy $v_{out} \gtrsim \sqrt{2GM_{BH}/r}$ for the gas to be unbound, where r is the radial distance from the black hole and $M_{BH}$ is the black hole mass. 
The minimum launch radius of the outflow calculated based on its velocity suggests that the outflow is launched at distances $\gtrapprox25 r_{s}$ during early times (X1-3, taking $v_{out}\sim-0.2c$), and at $\gtrapprox2500 r_{s}$ at late times (X7-8, taking $v_{out}\sim-0.02c$), where $r_{s}$ is the Schwarzschild radius. Even though this scenario can explain the diminishing strength of the UFO, it cannot account for its disappearance in X4 and X6 since we would expect the outflow to gradually decelerate across X1-8 if it were present.

\par The ionization parameter remains roughly constant over time, within uncertainties (Figure \ref{fig:ufo_timeevol}). By definition, the ionization parameter is given by $\xi = L/(\rho r_{l}^{2})$, where $L$ is the unabsorbed luminosity, $\rho$ is the outflow density, and $r_{l}$ is the launching radius. Since the luminosity decreases by approximately an order of magnitude and the outflow launching radius increases by a factor of 100, the overall density of the outflow should have dropped by almost five orders of magnitude in the $\sim$2 year time span between X1 to X8. We may extend the above definition of $\xi$ to derive an approximate expression for the density profile of the outflow as a function of the distance it traverses since its launch to be $\rho(r) = \rho_{0} (r/r_{l})^{- \alpha}$, where $\rho_{0}$ and $r_{l}$ give the launching density and launching radius of the outflow respectively, and $r$ gives the distance to which the outflow has expanded at a particular time. With the order-of-magnitude approximations outlined previously, we get $\alpha\approx$2.5, suggestive of the fact that the $\sim$10$^{5}$-factor decrease in the outflow density was caused by a rapid expansion and increase in volume of the outflow, and perhaps since the outflows at later times were launched from radii almost a hundred times the initial launch radius. With this scenario, it is difficult to explain the disappearance of outflows in X4 and X6, unless we assume the outflow to be ``patchy'', in which case, it is possible for an ionized blob within the outflow to move away from our line-of-sight and give the impression of a disappearing outflow. If we invoke a geometric change in the outflow by assuming that the outflow does not expand uniformly, but instead depends on the opening angle at which it is launched \citep[e.g., see][the higher the outflow velocity, the smaller is the opening angle]{2018MNRAS.479L..45P}, we may be able to explain the rapid decrease in the outflow density, and also the constant $\xi$. It is, however, still challenging to explain the complete outflow disappearance with such an expansion as the only factor.

\subsection{Outflow energetics}
\par We estimate the mass outflow rate for when outflows were observed using $\dot{M}_{out} = \Omega m_{p} N_{H} v_{out} r_{l}$ \citep[][hereafter denoted as $\dot{M}$]{Nardini2015}, where $\Omega$ is the solid angle of the launching outflow (taken to be $2\pi$, as a typical value for disk winds), $m_{p}$ is the proton mass, $N_{H}$ is the ionized column density, $v_{out}$ is the outflow velocity, and $r_{l}$ is the launching radius. For this calculation, we adopt the value for the black hole mass to be log($M_{BH}$/\(\textup{$M$}_\odot\))$=6.23$, as inferred by \cite{2017MNRAS.471.1694W} from host galaxy stellar velocity dispersion scaling. The mass outflow rate increases by more than one order in magnitude in X7+8 as compared to X1-5 (Appendix Figure \ref{fig:mass_rates}), possibly because of an increase in the ionized column density $N_{H}$ and the launching radius $r_{l}$. The total mass launched into the circumnuclear medium (CNM) can be estimated by a simple linear interpolation of the mass outflow rates over the duration of $\sim 700$ days, which comes to be $\approx(8.8\pm0.1)\times 10^{-3}$\(\textup{M}_\odot\). We can estimate the kinetic energy of the outflow using $\dot{K} = \dot{M} v_{out}^2/2$, from which we find that the net kinetic energy injected into the CNM in X1-8 was $\approx10^{48}$ ergs. Taking the host galaxy's (PGC043234) mass to be log($M_{BH}$/\(\textup{$M$}_\odot\))$ \approx 9.9$ \citep[computed using Equation 5,][]{Yao2023}, and a stellar dispersion velocity of $\sim$78 km/s \citep{2017MNRAS.471.1694W}, we find the approximate binding energy of the galactic bulge to be $4.8\times10^{56}$ ergs. With a TDE rate of $1.6\times10^{-5}$year$^{-1}$ for a system like ASASSN-14li showing constant blue colors \citep{Yao2023}, we find that the time taken for such a TDE to pump-in energy comparable to that of the galactic bulge would be $\sim$10$^{13}$ years. Consequently, ASASSN-14li's outflows are not dominant contributors to galactic feedback.

\begin{figure}
\includegraphics[width=\columnwidth]{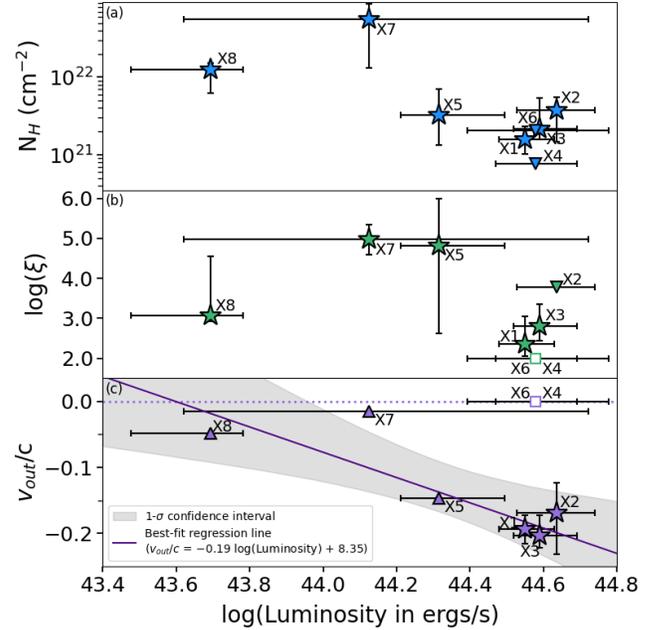}
\caption{Dependence of outflow parameters on bolometric luminosity of the source (13.6 eV to 13.6 keV) for X1-8 using the disk blackbody + UFO (XSTAR) model: (a) ionized column density ($N_{H}$ in cm$^{-2}$) as a function of the intrinsic luminosity for X1-8; (b) dependence of the ionization parameter ($\xi$, in logarithmic scale) on the intrinsic luminosity; and (c) luminosity dependence of the blue-shifted outflow velocity for each observation X1-8 in natural units ($v_{out}/c$). Star markers represent parameter values with errors, triangle markers with downward and upward arrows represent upper and lower limits for a particular parameter value respectively, and open square markers for X4 and X6 represent fixed parameter values for log($\xi$)=2 and $v_{out}=0c$ to obtain upper limits on the ionized columns in the respective cases. The bottom panel (c) shows the best-fit regression line for bolometric luminosity vs outflow velocity. The best-fit linear regression parameters here represent median values (since median is less-sensitive to outliers as compared to mean), and shaded grey region gives 1-$\sigma$ confidence intervals of the computed best-fit line.}
    \label{fig:ufo_lumin_dep} 
\end{figure}

\subsection{Radiation-driven outflows}
\par To test whether the outflows seen in ASASSN-14li were driven by radiation pressure, we checked for a correlation between the unabsorbed luminosity and outflow parameters (Figure \ref{fig:ufo_lumin_dep}). We focused on the luminosity correlation with outflow velocity, as we do not see an obvious correlation with the ionized column density, or the the ionization parameter. In a scenario where outflows are driven by radiation, we would expect the outflow velocity to decrease with time as the outward radiation pressure drops. While there appears to be a correlation between the outflow velocity and unabsorbed luminosity, we have too few data points to infer significant correlation and conclude that the outflows are radiation driven. Moreover, we will need an additional factor to explain the outflow disappearance in X4 and X6 if the outflows are driven by radiation.

\par An alternative explanation is the following: if the black hole powering ASASSN-14li has a mass of $\sim 10^{6}$\(\textup{M}_\odot\) and the disrupted star was solar-like, then the fallback rate (i.e., the rate of return of tidally destroyed material to the black hole) can be as high as $\sim$4\(\textup{M}_\odot\)yr$^{-1}$ (e.g., Figure 11 of \citealt{lawsmith20}, Figure 4 of \citealt{nixon21}). Assuming a radiative efficiency of 0.1, such that the accretion luminosity is $L = 0.1 \dot{M}c^2$, the peak accretion rate is super-Eddington by a factor of $\sim 150$ for an electron-scattering opacity of $\kappa = 0.34$ cm$^{2}$ g$^{-1}$ (valid for solar abundances). Letting the accretion luminosity scale as $L = L_{\rm peak}\left(t/T_{\rm peak}\right)^{-5/3}$, where $L_{\rm peak} = 150 L_{\rm Edd}$ and $T_{\rm peak} \simeq 25$ days (e.g., \citealt{golightly19, lawsmith20, bandopadhyay23}), 
the accretion luminosity should fall below the Eddington limit of the black hole after $\sim 500$ days post-maximum, in agreement with the observed Eddington ratios in ASASSN-14li (Figure \ref{fig:tempevol}). If the star was completely destroyed or nearly completely destroyed during its initial tidal interaction, which is required to produce a fallback rate of $\sim 4 M_{\odot}$ yr$^{-1}$, roughly half of the original mass of the star would be accreted by the time the accretion rate becomes sub-Eddington.
 
\par The existence of super-Eddington accretion in ASASSN-14li is consistent with the formation of a wind that is powered radiatively, as discussed in the preceding paragraphs. Additionally, however, the super-Eddington nature of the accretion implies that the surrounding accretion flow should be geometrically thick, and for the relatively large temperatures and low densities of the gas, predominantly radiation-pressure supported. We would then expect the pressure of the surrounding ``envelope'' (quite distinct from a standard thin disk) to collimate the outflow, either in a viscous-like manner that is mediated by the interaction of the radiation in the surrounding envelope and the outflow (as described in more detail in the next paragraph), or through the formation of oblique shocks \citep{bromberg07, kohler12}. This collimation could confine the fast-moving gas and the absorption signature to a sufficiently narrow solid angle that it would be -- unless we are fortuitously aligned along the axis of the outflow -- out of our line of sight. 

\par \cite{coughlin20} showed that structured outflows, with the bulk of the mass near the angular extremities and the fastest-moving and low-density material along the axis, naturally arise from the viscous-like interaction between a fast-moving jet and a surrounding, optically thick and radiation-supported envelope. The ability of the envelope to collimate the outflow, however, depends on the radial pressure gradient of the envelope, with pressure profiles shallower than $\propto r^{-2}$ (i.e., if the pressure $p$ varies with spherical radius $r$ as $p \propto r^{-n}$, then $n \le 2$) able to successfully confine the outflow and causing the streamlines to asymptotically approach the axis in angle. For typical TDE parameters, the surrounding accretion flow modeled with the ZEro-BeRnoulli Accretion (ZEBRA) solution \citep{coughlin14} predicts that the pressure profile should vary with spherical radius as $p \propto r^{-n}$, where $n$ ranges from $\sim 3/2$ -- $3$, and declines with time. 

\par Thus, it seems possible that the supercritical fallback rate in ASASSN-14li resulted in a radiation-driven outflow and a surrounding envelope with a pressure profile close to the limiting value for collimation of $p \propto r^{-2}$. The power-law index of the ZEBRA envelope is approximately related to the ratio of the total disk mass to the total angular momentum \citep[see Equation 19 of][]{coughlin14}, and so, if the point of closest approach of the disrupted star was particularly small, as required to completely destroy a solar-like and radiative star \citep{guillochon13, mainetti17}, it would drive the pressure power-law index to larger values. When the pressure profile is shallower than $p \propto r^{-2}$, the fast-moving material is confined to a sufficiently small opening angle that the optically thick envelope obscures our line of sight, rendering the outflow and the absorption signature unobservable. As the pressure profile steepens past $p \propto r^{-2}$, the outflow de-collimates, resulting in the formation of a laterally propagating shock as the outflow loses causal contact and the ejection of a larger amount of mass. This widening of the outflow and ejection of mass simultaneously places fast-moving material along our line of sight, thus enabling the formation of the absorption signature, and reduces the pressure profile (by removing low-angular momentum material and thus reducing the ratio of the total disk mass to angular momentum) to the point that the outflow is once again collimated. In this way, the disk can periodically pressure confine and de-confine the outflow and modulate the strength and appearance of the absorption signature, thereby explaining the transient nature of the outflow. 

\subsection{Alternative models}
\par Recent general-relativistic magneto-hydrodynamic (GR-MHD) simulations have suggested that a stellar-sized object in the vicinity of a supermassive black hole (SMBH) can perturb and punch through the accretion disk, repeatedly driving outflows along the magnetic rotation axis of the black hole \citep{Sukova2021}. If we adopt this setup in ASASSN-14li, we can explain the transient nature of outflows, including the outflow's appearance and disappearance. However, without a clear periodicity and phase-resolved X-ray spectra over that period, it is challenging to verify this model with the data used in this work.

\par Finally, as a potential alternative model, it has been suggested that the late-time X-ray flares often detected from gamma-ray bursts (GRBs; e.g., \citealt{burrows05, margutti11}) could be associated with the accumulation of magnetic flux onto the black hole (in the case of a collapsar/long-GRB model; \citealt{woosley93, macfadyen99}) or neutron star (in the case of a compact-object merger/short-GRB model; \citealt{paczynski86, goodman86}), as accretion could lead to the formation of a split-monopole-like field \citep{proga06}. In a TDE, the magnetic field associated with the original star should be predominantly in the direction of the returning debris stream (i.e., owing to magnetic flux conservation, the component of the magnetic field perpendicular to the stream axis declines more rapidly than the component along the stream; \citealt{bonnerot17, guillochon17, coughlin21}), resulting in a largely toroidal field that is not conducive to the formation of a fast outflow \citep{blandford82}. However, it may be possible to turbulently amplify the radial component of the field during the accretion process (through the magnetorotational instability; \citealt{balbus91}), leading to late-time magnetorotational energy extraction and the formation of an episodic outflow, consistent with the observations reported here.

\section{Summary}
\label{section:conclusion}
\par In summary, by tracking the UFO from the ``Rosetta Stone'' of TDEs, ASASSN-14li, we report the following:
\begin{enumerate}
    \item Evolution of the absorption feature detected in the X-rays is complex and intermittent, with the feature appearing and disappearing up to 2 years following the TDE discovery. The velocity associated with the absorption signature also declines with time and would not qualify as an ``ultra-fast'' outflow by 2 years, but we interpret the signature as arising from the same physical origin, i.e., an outflow that is more slowly moving at later times. Observations recorded $> 2$ years post optical discovery suggest that the feature was absent during late times.
    \item Such repetitive TDE outflow behavior is inefficient in pumping energy into its host galaxy and is therefore unlikely to contribute significantly towards galaxy feedback.
    \item ASASSN-14li does not show evidence for a corona out to 4.5 years, which is expected given the consistently high Eddington ratios over this timeline. This could be indicative of a complete absence of the corona, or, its weak existence.
\end{enumerate}

\par Further, we have explored various models in order to explain the above results. While the decrease in strength of the outflow can be attributed to larger launching radii and a rapid decline in the outflow density at late times, it cannot explain the transient outflows observed in ASASSN-14li. Mapping the association between the bolometric luminosity of the source and the outflow velocity points towards a potential correlation, signifying a scenario involving radiatively driven winds or outflows. However, to explain the recurring outflow behavior, it becomes necessary to invoke the notion of structured outflows that are influenced by the ability of the surrounding the accretion flow (well-modeled with the ZEBRA solution for pressure profile), to collimate the outflow. Then, the observed absorption signature is ``mirrored'' in the X-ray spectrum when the outflow is collimated sufficiently into our line of sight. We briefly touch-upon two other models that could address the presence (and absence) of outflows in the source -- one, involving a stellar-sized perturber in orbit around the black hole, driving succesive outflows along our line of sight; and two, a model analogous to those discussed in the context of GRBs, involving the accumulation of magnetic flux onto the black hole that can eventually lead to the launch of late-time outflows. 

\par The detection of transient outflows from the TDE ASASSN-14li necessitates further investigation to determine whether they are exclusive to this system or if they can occur in other systems. Should these outflows prove to be common, it will become crucial to identify and compare the mechanisms governing the outflow behavior in different systems, which would provide fresh insights into the interplay between accretion dynamics and the launch of outflows.

\section{Acknowledgments}
The authors thank the anonymous referees for the helpful comments and feedback on the manuscript. D.R.P was partly supported by NASA/XMM grant 80NSSC21K0835. E.R.C.~acknowledges support from the National Science Foundation through grant AST-2006684, and a Ralph E.~Powe Junior Faculty Enhancement Award through the Oakridge Associated Universities. M.G. is supported by NASA XMM-Newton grants 80NSS23K0621 and 80NSSC22K0571.

\bibliography{refs.bib}{}
\bibliographystyle{aasjournal}

\appendix
\label{appendix}
\counterwithin{figure}{section}
\counterwithin{table}{section}
\section{Pile-up correction}
\label{section:pileup}
\par Pile-up is pronounced in sources with high X-ray brightness. This can lead to the detection of lower photon counts by the instrument and a shift toward higher energies in the X-ray spectrum \citep{2015A&A...581A.104J}. In this section, we quantify pile-up in the EPIC-pn observations used in our work and describe the procedures adopted to mitigate it.

\par We first assessed the proportion of single (s) and double (d) pixel pn-events that showed agreement with the predicted pattern distribution by using the {\tt epatplot} tool in SAS. Since events with observed-to-model fractions $s<1$ (for single pixel events) and $d>1$ (for double pixel events) indicate the presence of pile-up, we ensured that both these ratios were consistent with unity within statistical uncertainty. We started by excising a circular region of ${4"}$ from the center of the source and systematically increased the exclusion radius by units of ${1"}$, each time checking for the overlap and ratios, until we obtained the appropriate $s$ and $d$ ratios close to unity. Table \ref{table:2} shows the final exclusion radii for each observation, and Figures \ref{fig:epat_w_pileup_corr}-{\ref{fig:epat_no_pileup_corr} show the corresponding pattern distribution diagrams before and after accounting for pile-up in each observation. The effects of pile-up are significant only in X1-X5 and we therefore considered excluded regions only for these observations. We did not exclude any regions for X6-X13 since pile-up is negligible in these observations.

\begin{flushleft}
\begin{table*}[h!]
\begin{tabular}{|c||c|c|c|c|} 
 \hline
   & \multicolumn{2}{c|}{Exclusion Radius} & Initial count rate (cps) & Extracted source-region count rate (cps)\\ [0.5ex]
  & physical units & arcsec & & \\
 \hline\hline
 X1 & 220 & 11 & 10.70 $\pm$ 0.04 & 3.50 $\pm$ 0.02\\ 
 \hline
 X2 & 280 & 14 & 10.40 $\pm$ 0.03 & 1.76 $\pm$ 0.02\\
 \hline
 X3 & 300 & 15 & 8.91 $\pm$ 0.02 & 1.70 $\pm$ 0.01\\
 \hline
 X4 & 220 & 11 & 2.25 $\pm$ 0.01 & 0.83 $\pm$ 0.01\\
 \hline 
 X5 & 200 & 10 & 1.62 $\pm$ 0.01 & 0.65 $\pm$ 0.01\\ 
 \hline
 X6 & 0 & 0 & 1.80 $\pm$ 0.02 & -\\   
 \hline
 X7 & 0 & 0 & 0.68 $\pm$ 0.01 & -\\   
 \hline
 X8 & 0 & 0 & 0.35 $\pm$ 0.01 & -\\   
 \hline
 X9 & 0 & 0 & 0.20 $\pm$ 0.01 & -\\   
 \hline
 X10 & 0 & 0 & 0.12 $\pm$ 0.01 & -\\   
 \hline 
 X11 & 0 & 0 & 0.06 $\pm$ 0.01 & -\\   
 \hline
 X12 & 0 & 0 & 0.05 $\pm$ 0.01 & -\\   
 \hline
 X13 & 0 & 0 & 0.04 $\pm$ 0.01 & -\\  [1ex]
 \hline
\end{tabular}
\caption{Inner exclusion radius, and net count rates between 0.25 - 1keV in each observation before and after accounting for pile-up; pile-up is negligible in observations X6-13.}
\label{table:2}
\end{table*}
\end{flushleft}

\begin{figure}[ht]
\centering
    \includegraphics[scale=1.1]{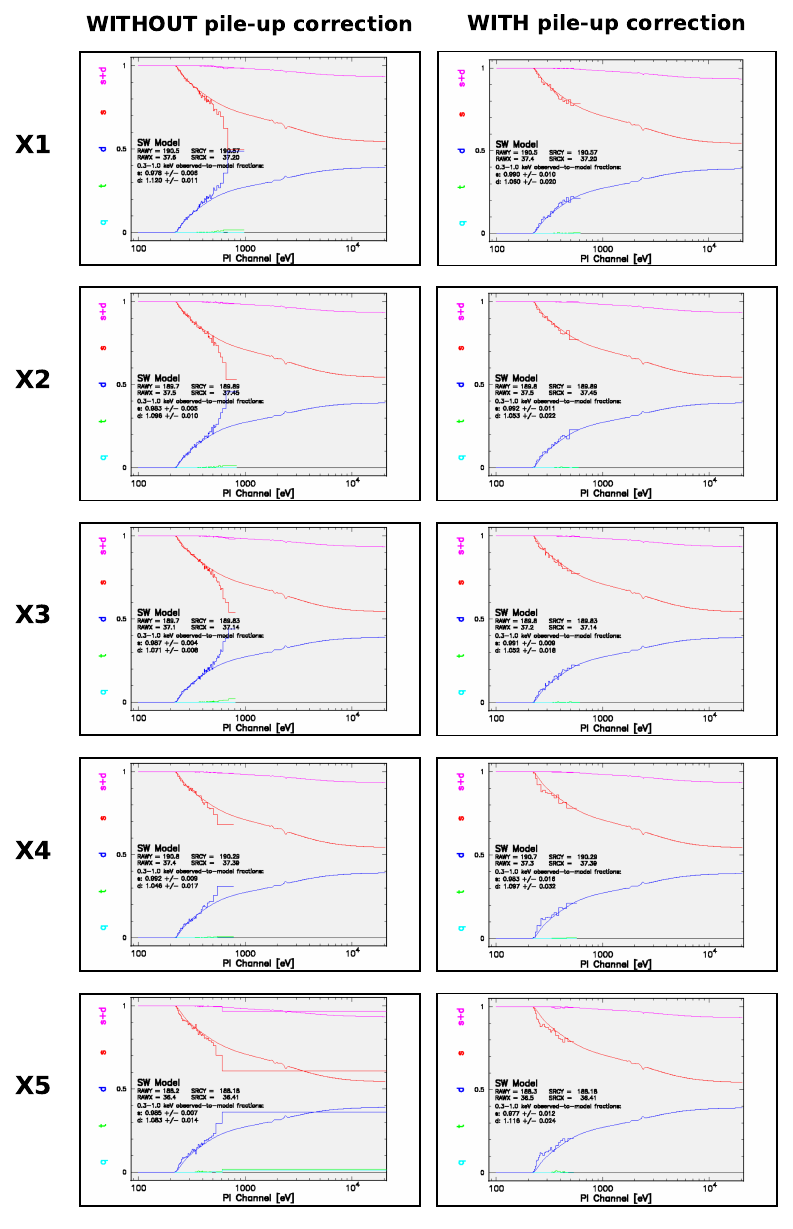} 
    \caption{Pattern distributions using the XMM-SAS {\tt epatplot} tool for observations X1-5 without pile up correction (left), and after excising an inner circular region (right).}
    \label{fig:epat_w_pileup_corr}
\end{figure}

\begin{figure}[ht]
\centering
    \includegraphics[scale=1.1]{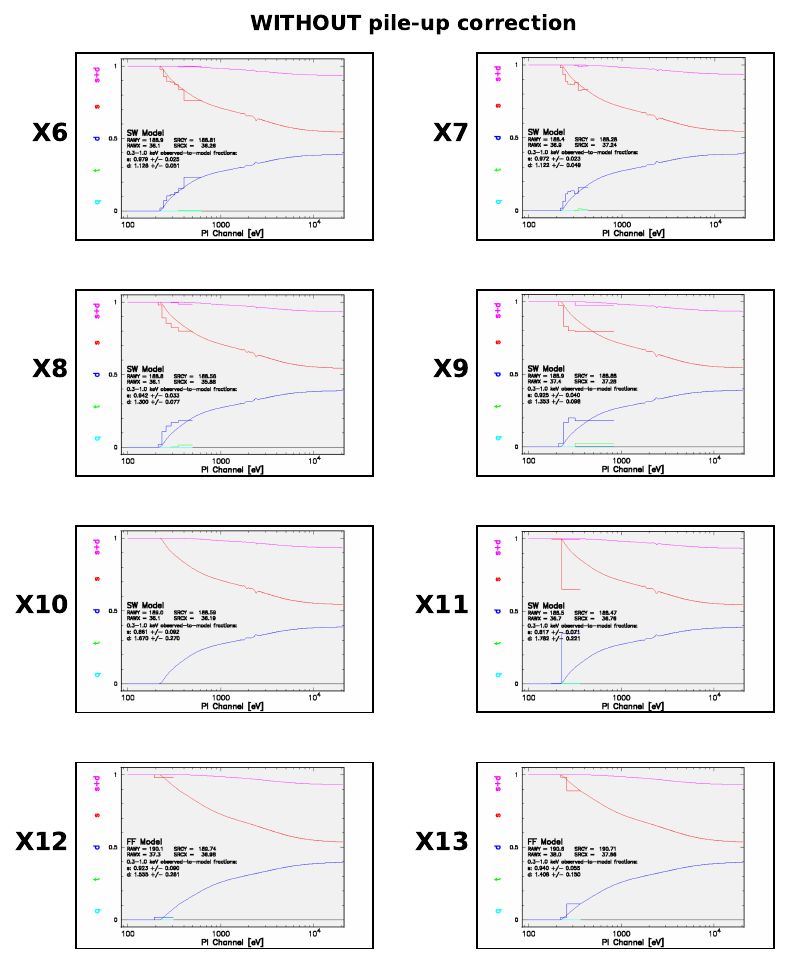} 
    \caption{Pattern distributions obtained using the XMM-SAS {\tt epatplot} tool for observations X6 to X13 where the effects of pile-up are negligible.}
    \label{fig:epat_no_pileup_corr}
\end{figure}
\clearpage

\section{Statistical tests}
\label{section:stats}

\subsection{Test I: Akaike Information Criterion (AIC)}
\label{subsection:aic}
$\Delta$AIC \citep{aic} for each case was calculated as follows:
$$ \text{C-stat}_{\tt XSTAR+diskbb} - \text{C-stat}_{\tt diskbb} + (2\times \text{d.o.f}),$$ 
\\where d.o.f = 3, as XSTAR adds 3 additional degrees of freedom (ionized hydrogen column density, $N_{H}$; the ionization parameter, log($\xi$); and the line-of-sight velocity, $v_{out}$) to the model. In general:
\begin{enumerate}[i]
    \item $\Delta$AIC $< -5$ indicates a preference for the complex model
    \item $\Delta$AIC $< -10$ indicates a strong preference for the complex model
\end{enumerate}

\subsection{Test II: {\tt simftest}}}
\label{subsection:simf}
We generated a set of 10,000 simulated datasets (for X1-8 individually) to obtain the probability that the absorption feature results from statistical noise, using the models described in Section \hyperref[subsection:diskufo]{3.2}. Results from {\tt simftest} can be visualized as histograms that give the fraction of simulations where $\Delta$C-statistic is lesser than the observed $\Delta$C-statistic (red line) for that observation, as illustrated in Figure \ref{fig:simftest}. This fraction then represents the spectra that have a statistically significant feature, and are not contaminated by noise -- X1, X2, X3, X5, X7, and X8 in our case. 

\begin{table*}[ht]
\centering
\begin{tabular}{|c|c|c|}
\hline
    Observation & $\Delta$AIC & {\tt simftest} probability\\
\hline \hline
    X1 & $-23.0$ & $<0.0001$\\ 
    \hline
    X2 & $-14.6$ & $<0.0001$\\
    \hline
    X3 & $-40.2$ & $<0.0001$\\
    \hline
    X4 & $5.8$ & $<0.3246$\\
    \hline
    X5 & $-7.1$ & $<0.0014$\\
    \hline
    X6 & $4.4$ & $<0.1088$\\
    \hline
    X7 & $-27.9$ & $<0.0001$\\
    \hline
    X8 & $-4.3$ & $<0.0001$\\
    \hline
\end{tabular}
\label{table:stat}
\caption{Results for AIC and {\tt simftest}. A description of each of these tests is provided in the main text (Section \ref{subsection:tests}) and above.}
\end{table*}

\begin{figure}[ht]
    \centering
        \includegraphics[scale=0.45]{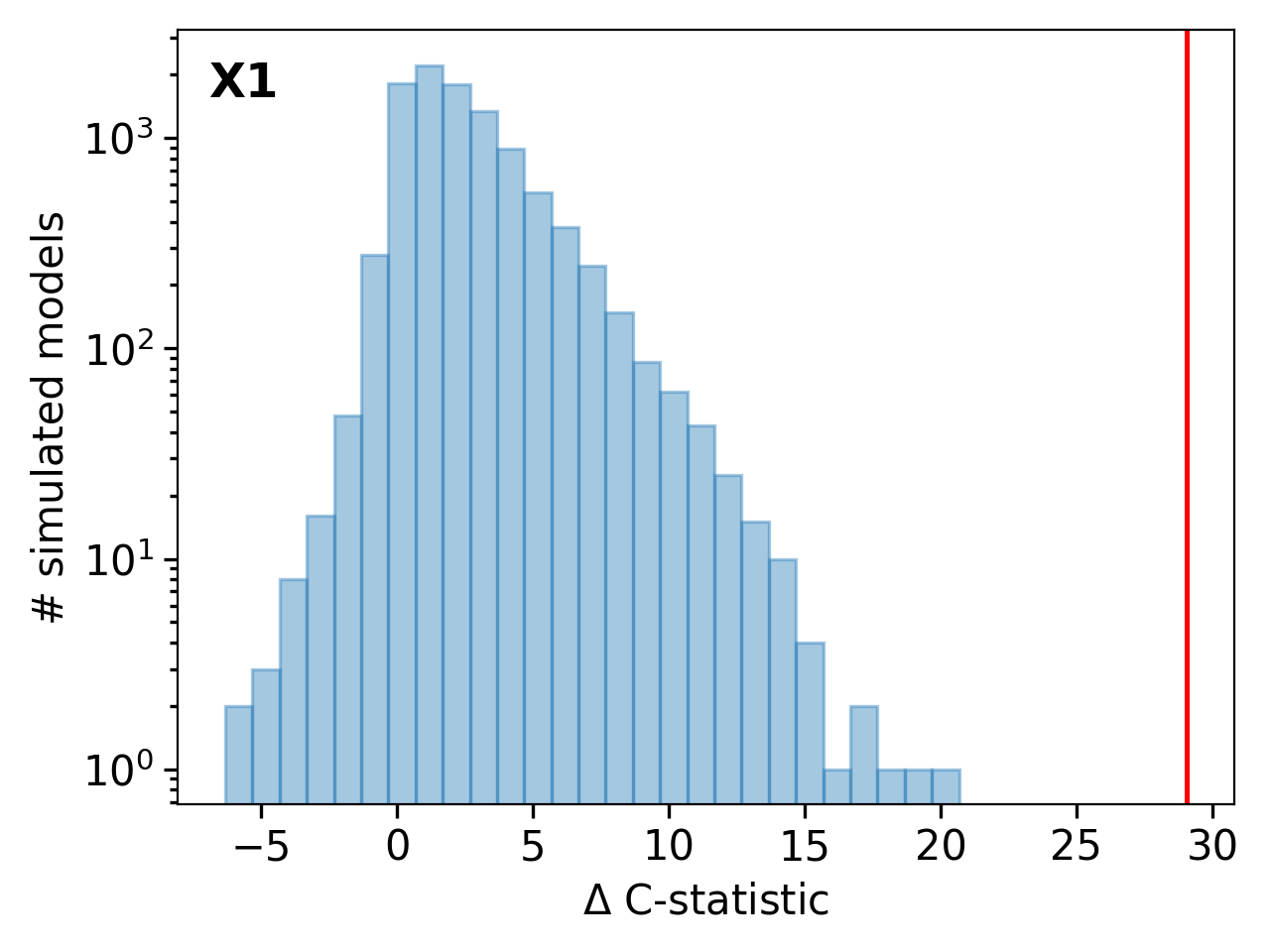}
        \includegraphics[scale=0.45]{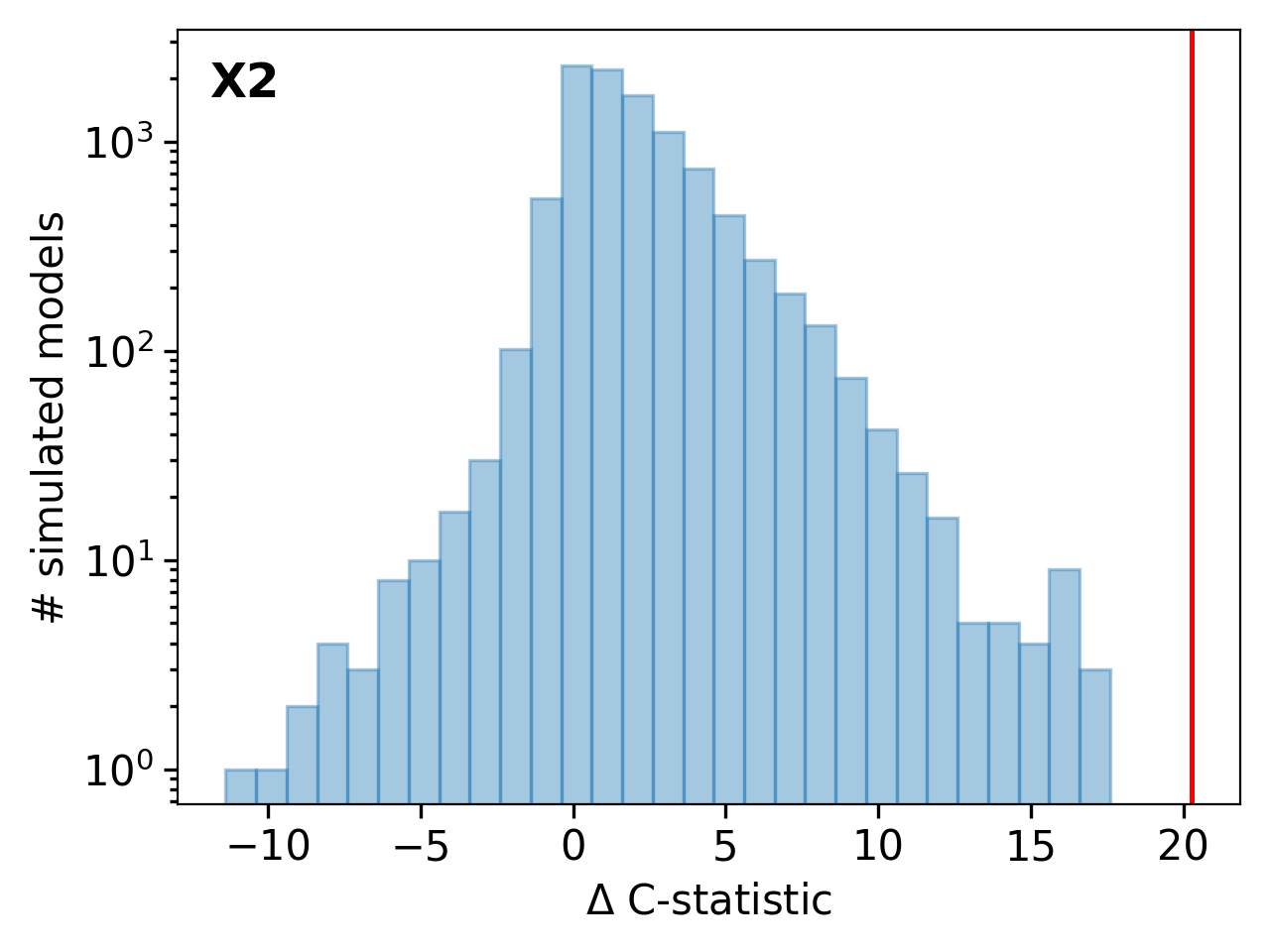}\\
        \includegraphics[scale=0.45]{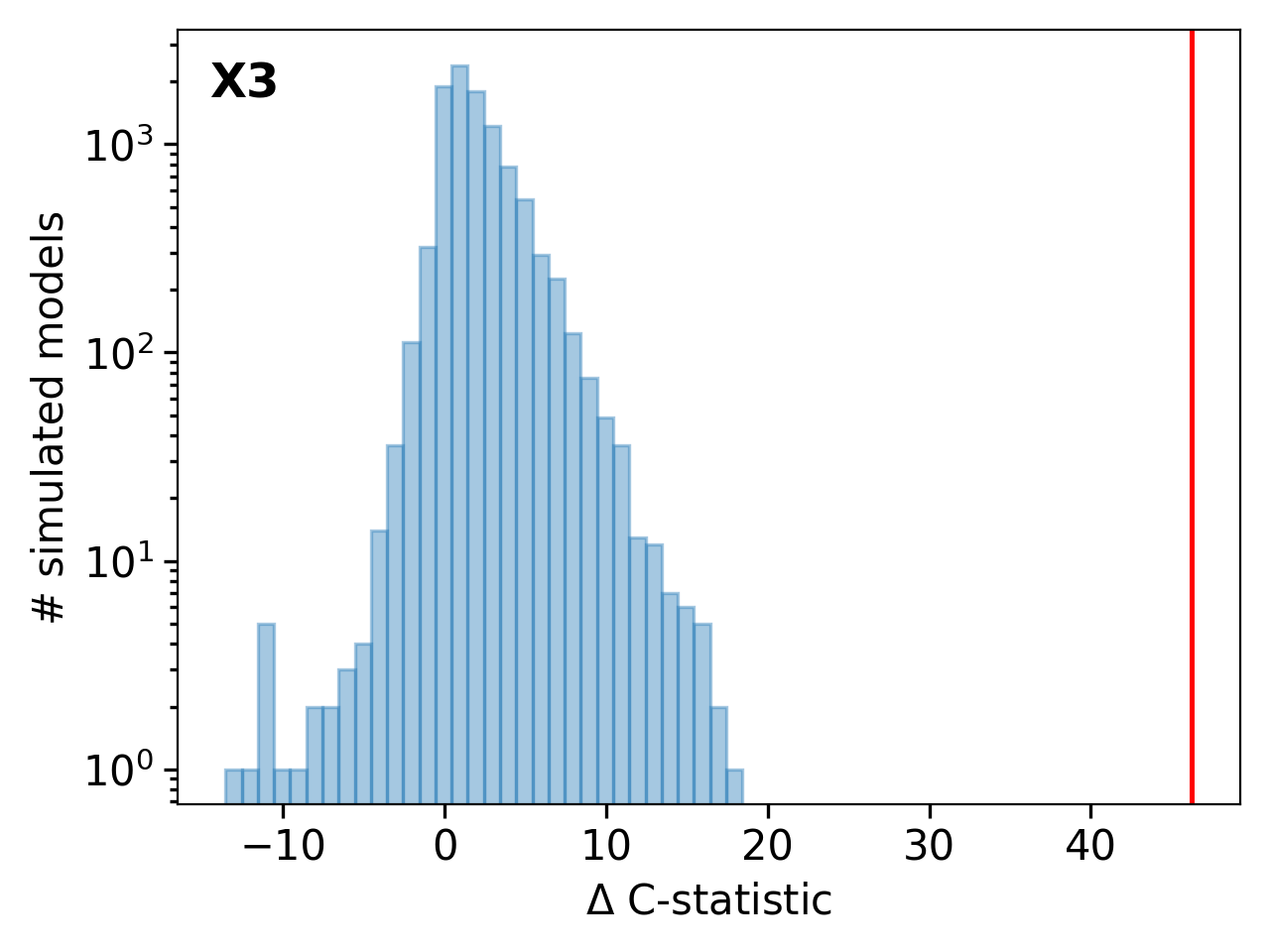}
        \includegraphics[scale=0.45]{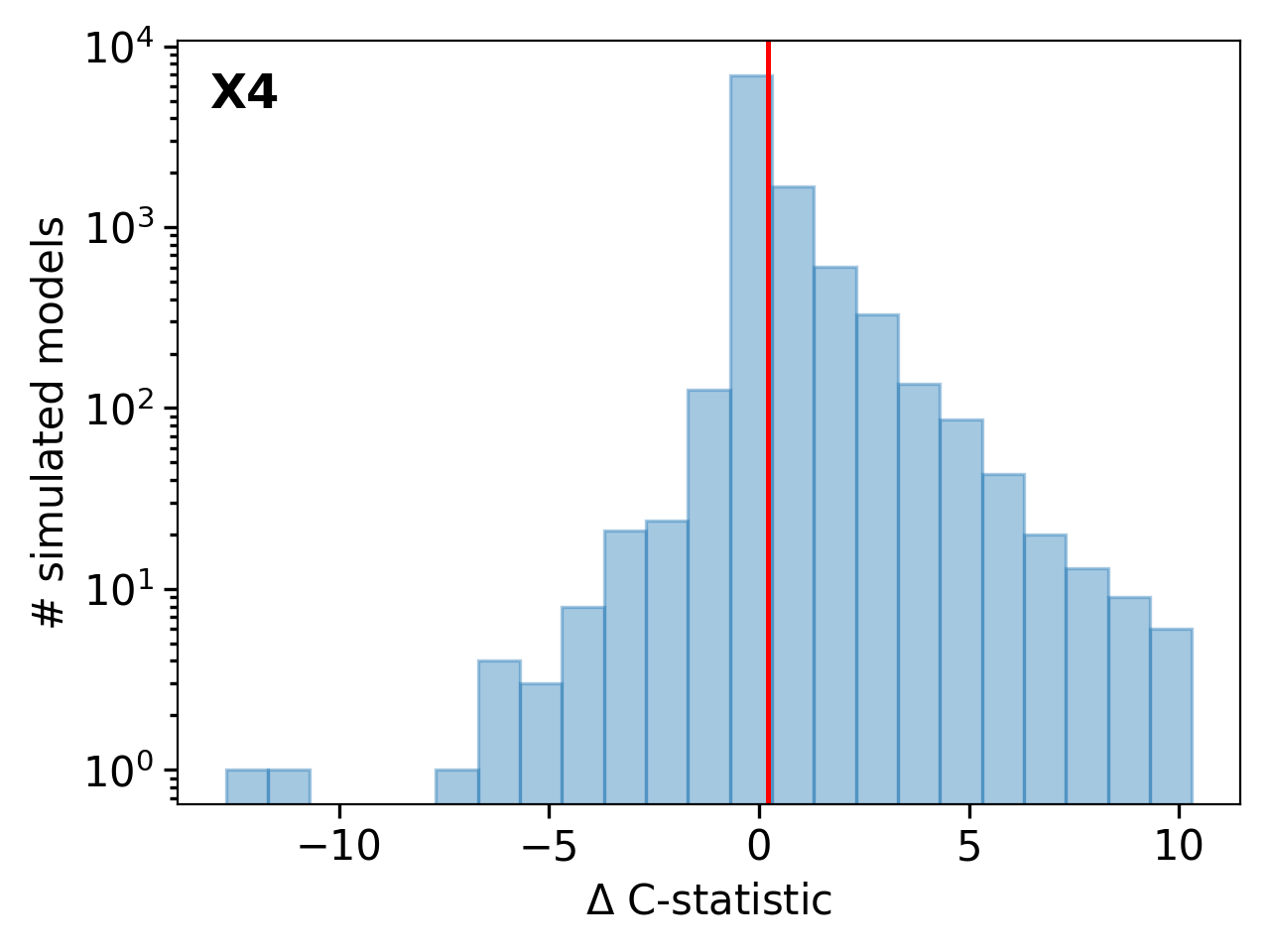}\\
        \includegraphics[scale=0.45]{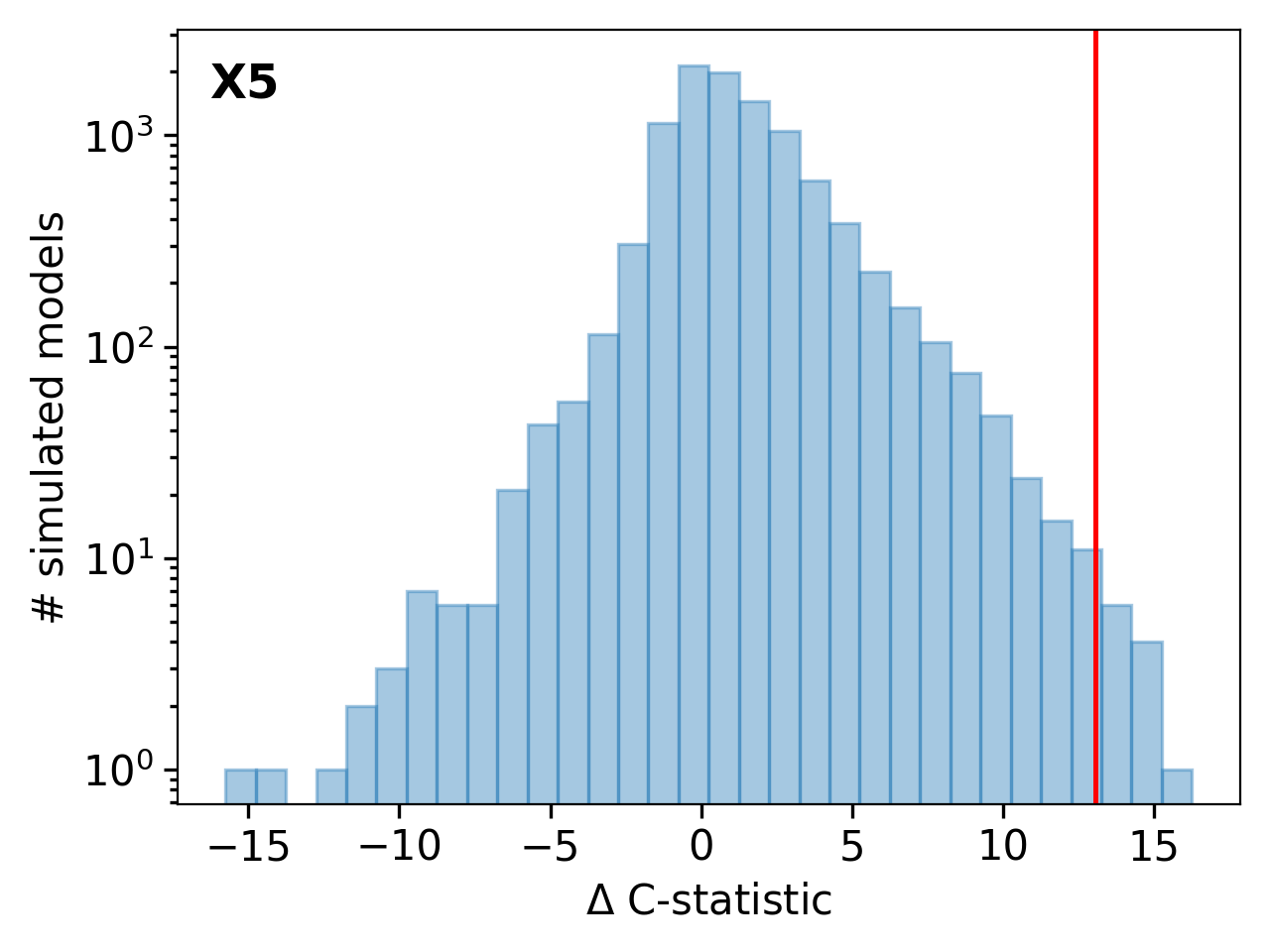}
        \includegraphics[scale=0.45]{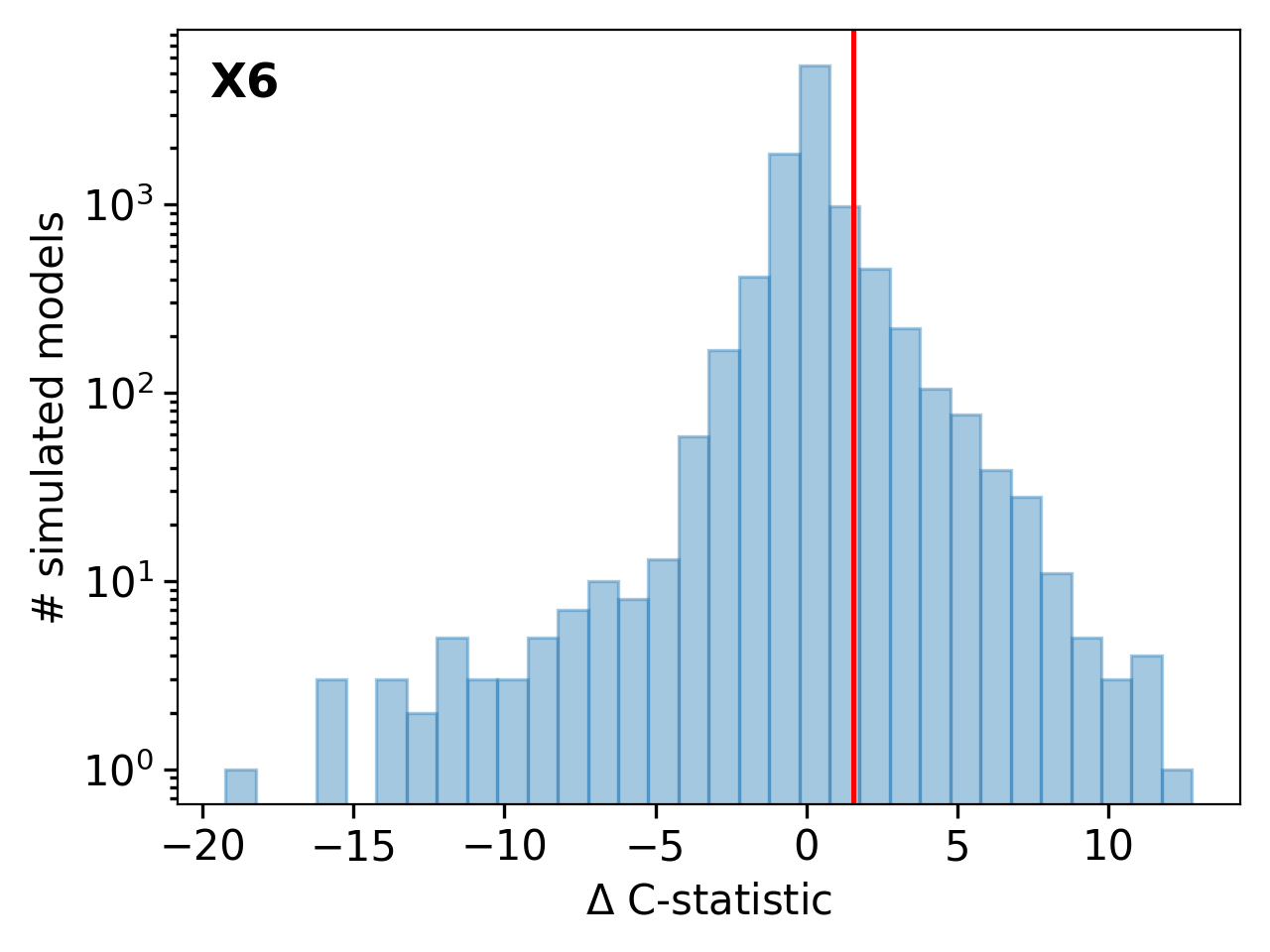}\\
        \includegraphics[scale=0.45]{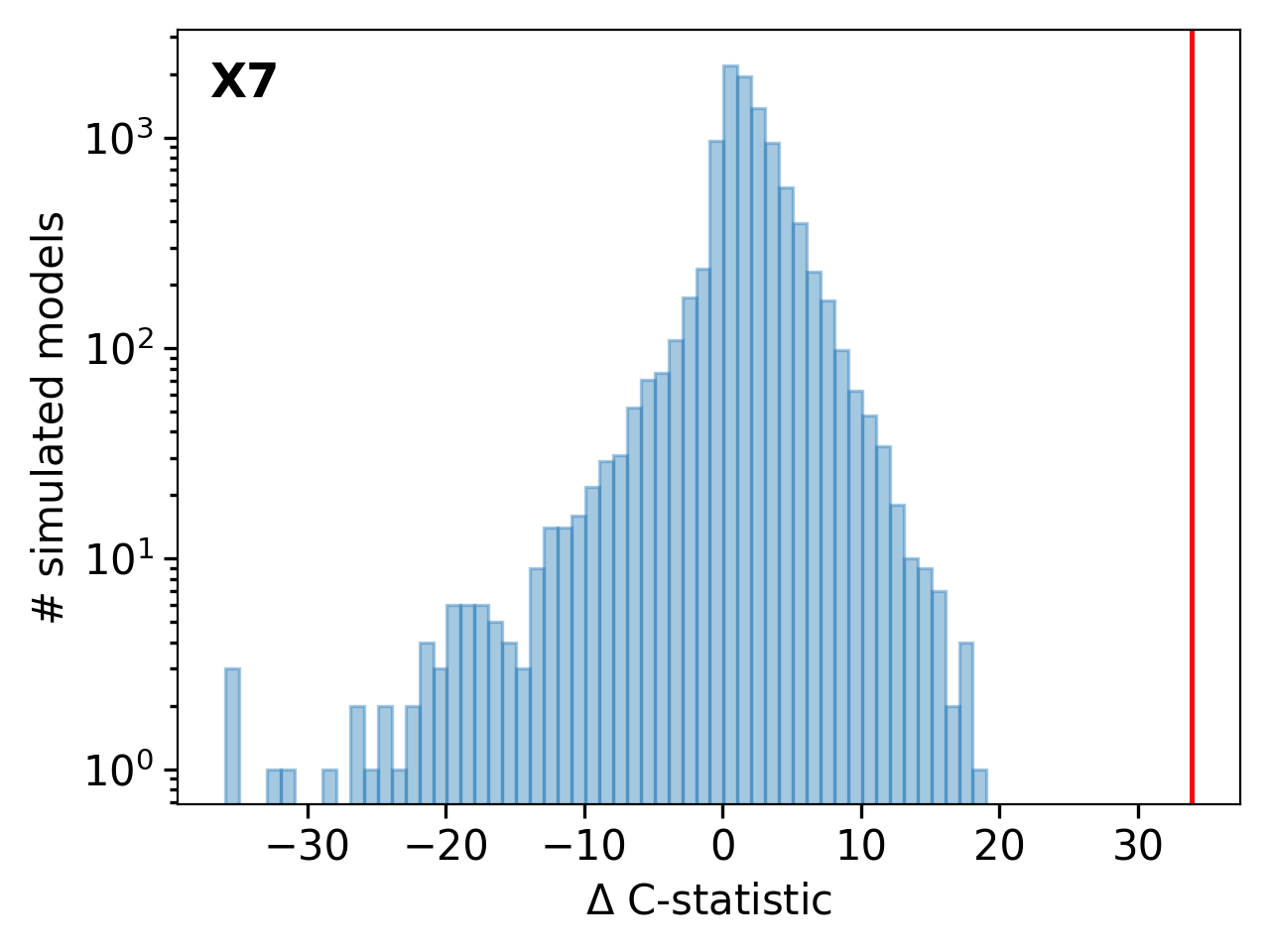}
        \includegraphics[scale=0.45]{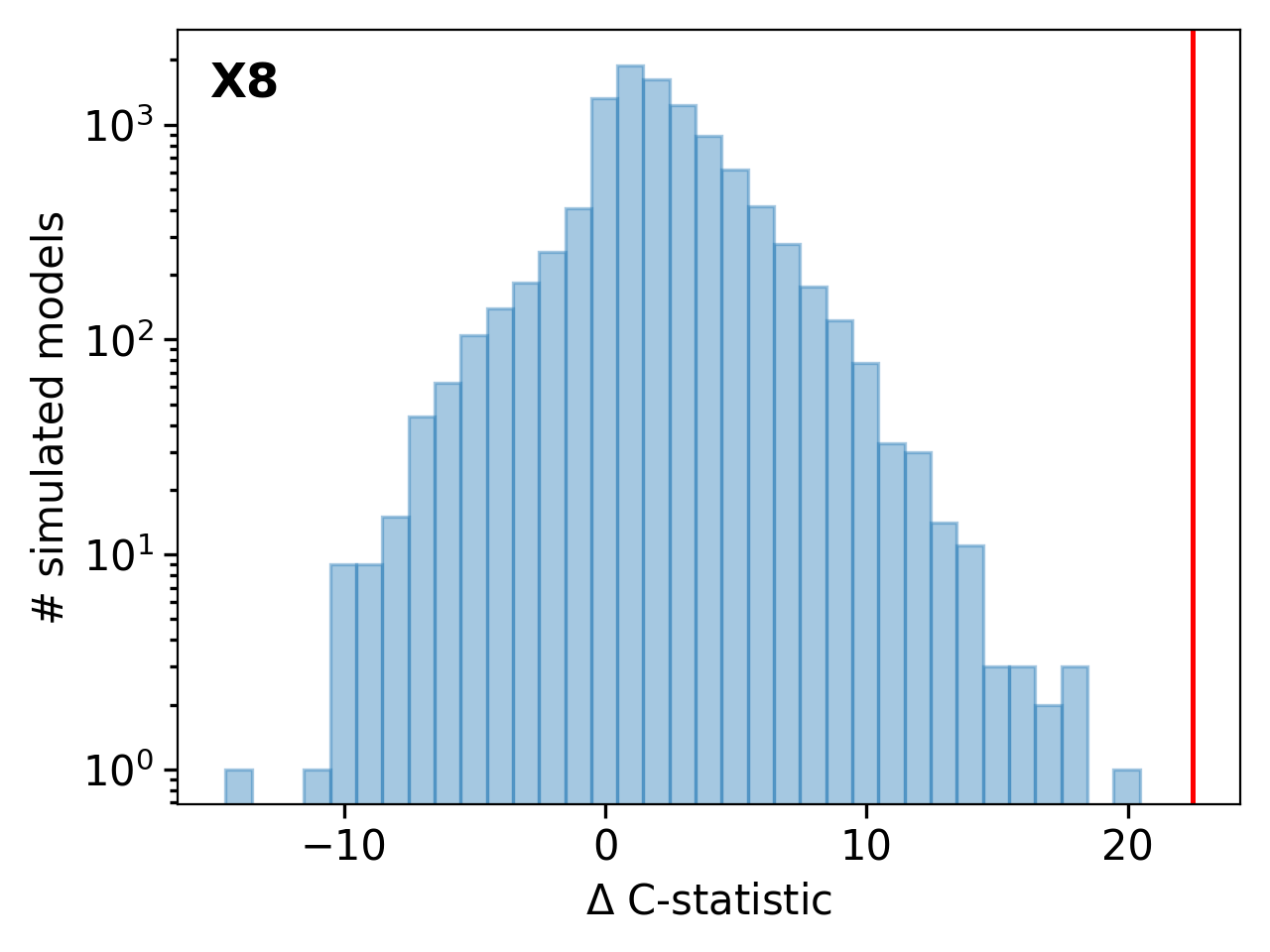}
    \caption{Results from {\tt simftest} discussed in Appendix \ref{subsection:simf}.}
    \label{fig:simftest}
\end{figure}

\section{Confidence contours with {\tt steppar} tool in XSPEC}
\label{subsection:steppar}
Confidence contours for the XSTAR parameters in observations with a statistically significant absorption feature have been shown in the Figures C1 - C5  for a visualization of the parameter space. We plot the 1-$\sigma$ (green), 2-$\sigma$ (blue), and 3-$\sigma$ (purple) contours calculated based on the C-statistic obtained by running {\tt steppar} for each of the following pairs of parameters -- ionized column density and ionization parameter, ionized column density and outflow velocity, ionization parameter and outflow velocity. The black cross-markers show the best-fit values.

\begin{figure}[ht]
\centering
\includegraphics[scale=0.62]{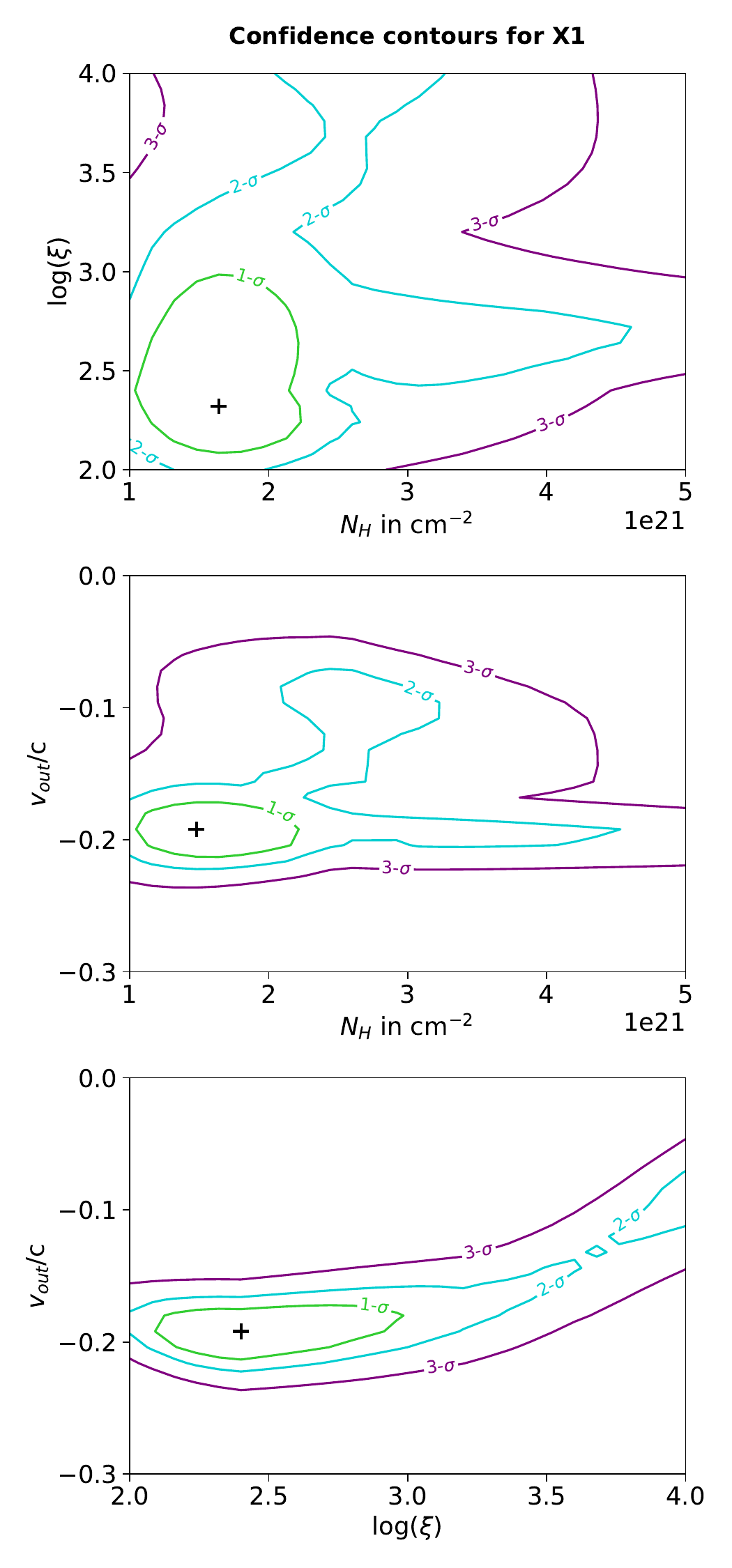} 
\caption{Confidence contours for X1 obtained using the {\tt steppar} tool (discussed in Appendix \ref{subsection:steppar}). }
\end{figure}

\begin{figure}[ht]
\centering
    \includegraphics[scale=0.62]{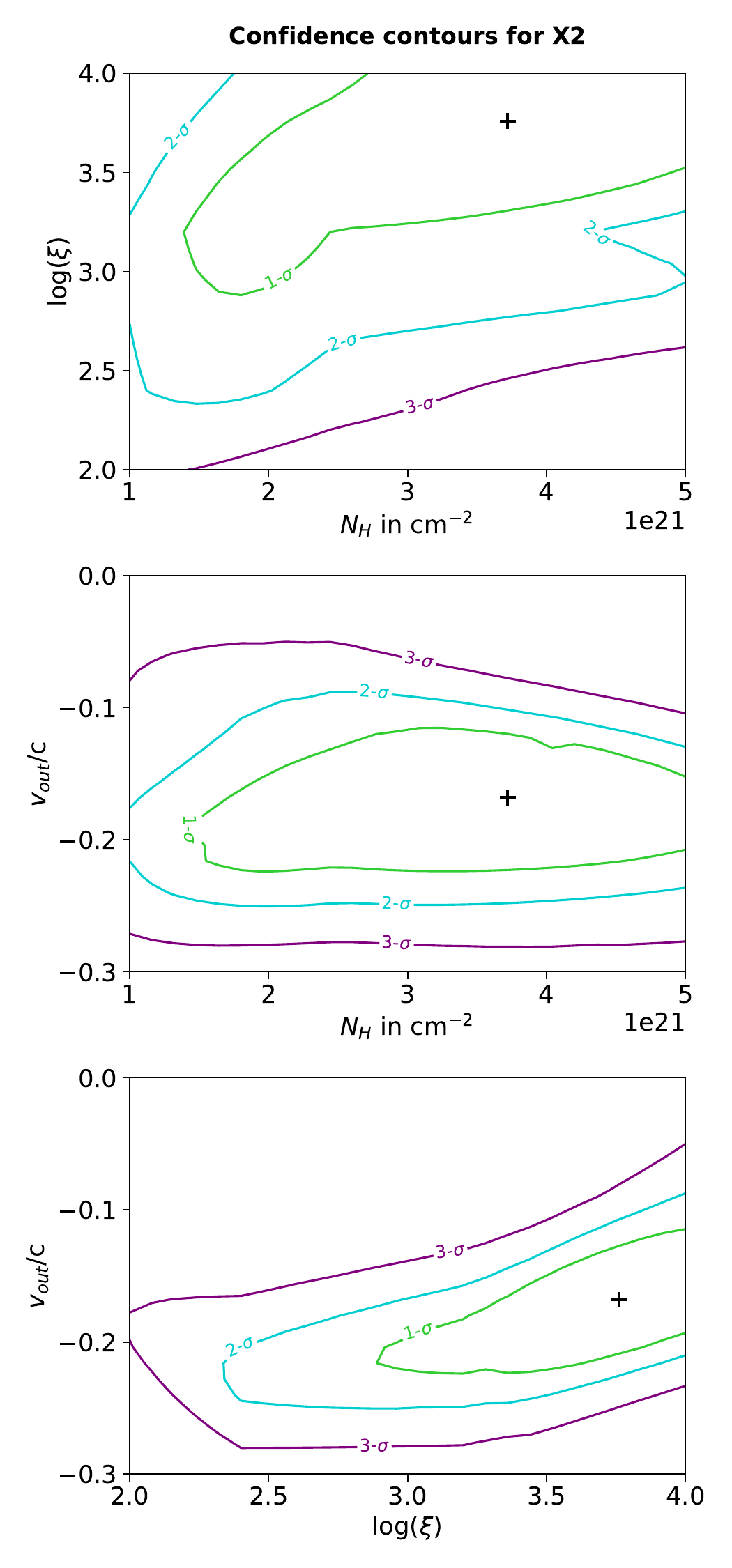}
    \caption{Confidence contours for X2 obtained using the {\tt steppar} tool (discussed in Appendix \ref{subsection:steppar}). }
\end{figure}

\begin{figure}[ht]
\centering
    \includegraphics[scale=0.6]{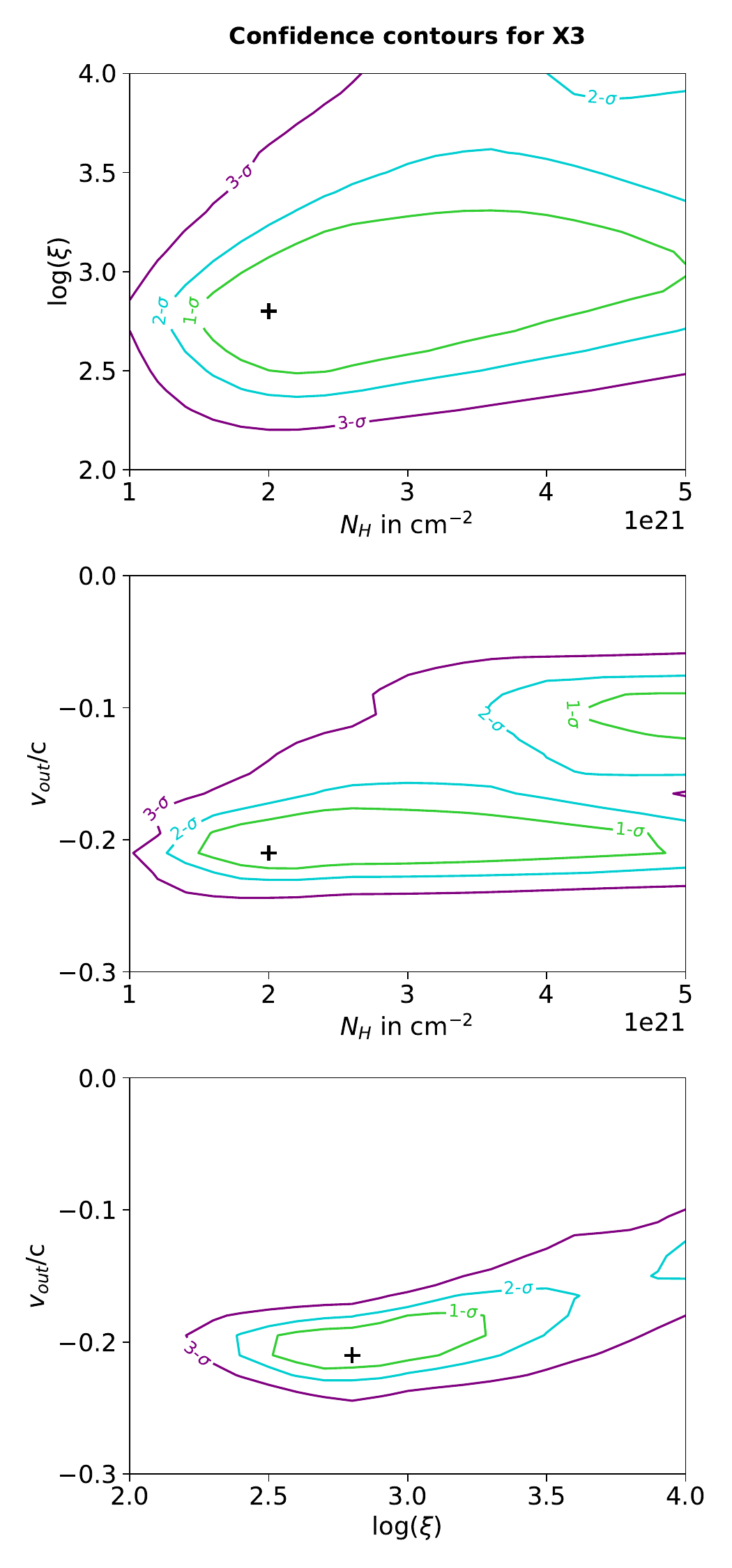} 
    \caption{Confidence contours for X3 obtained using the {\tt steppar} tool (discussed in Appendix \ref{subsection:steppar}). }
\end{figure}

\begin{figure}[ht]
\centering
    \includegraphics[scale=0.60]{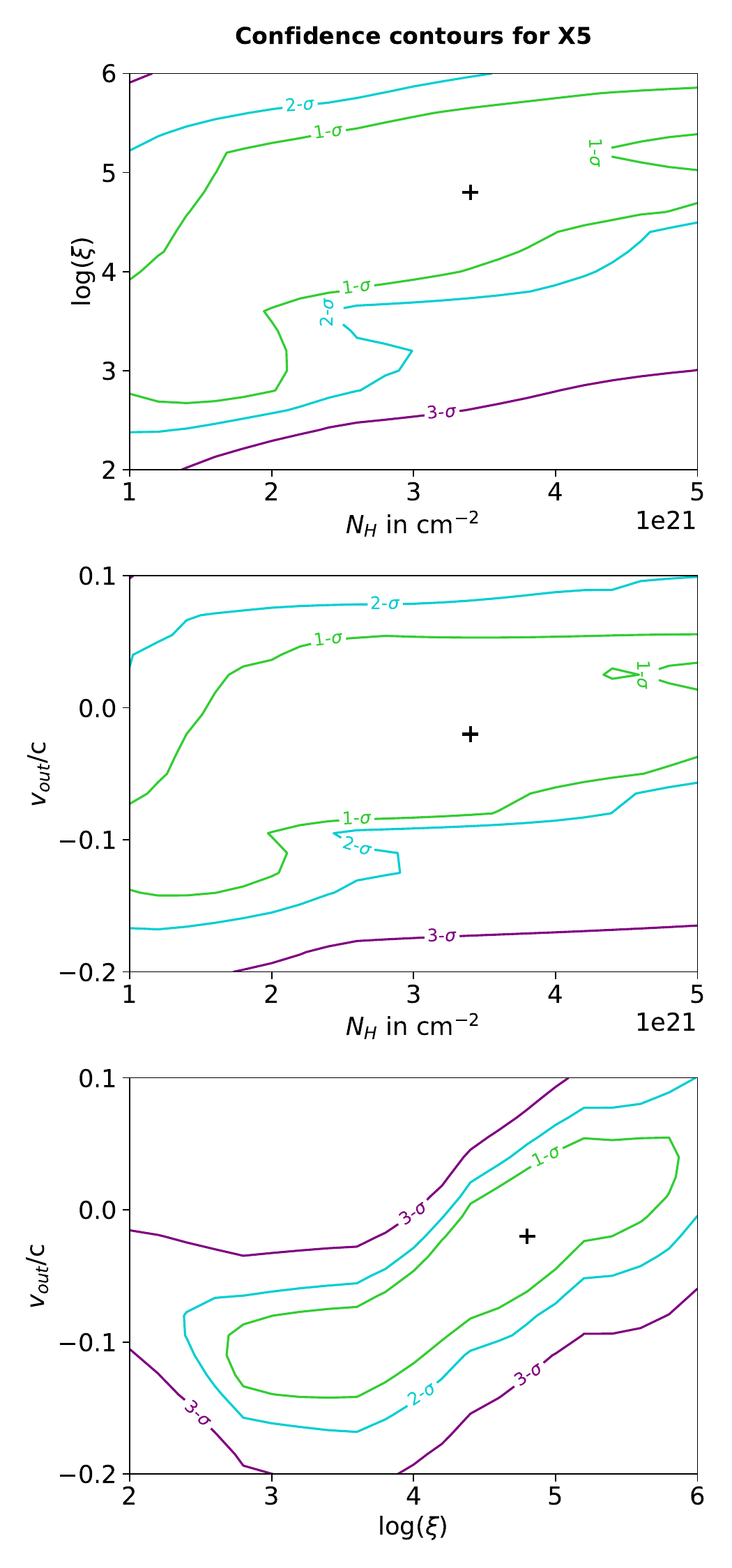} 
    \caption{Confidence contours for X5 obtained using the {\tt steppar} tool (discussed in Appendix \ref{subsection:steppar}). }
\end{figure}

\begin{figure}[ht]
\centering
    \includegraphics[scale=0.6]{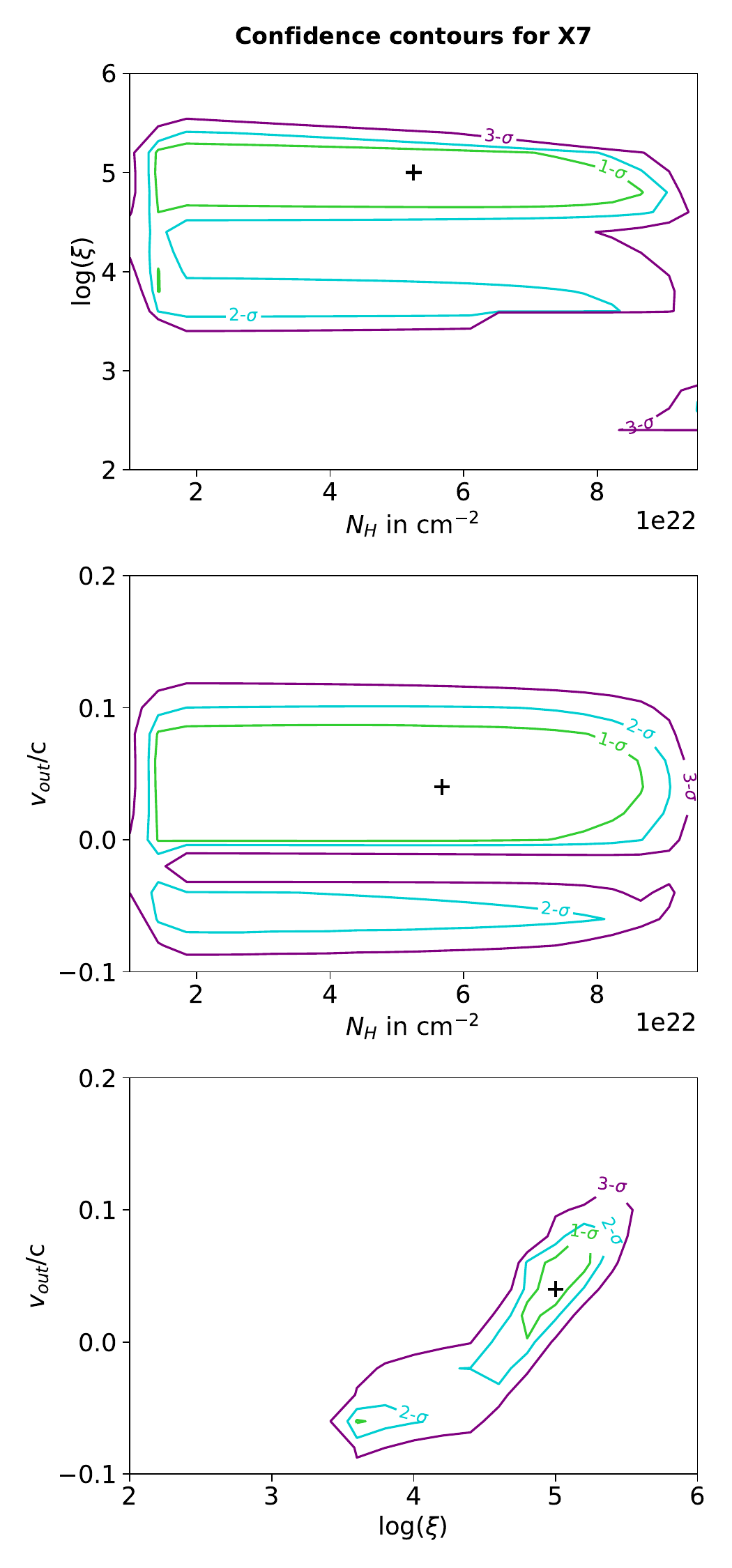} 
    \caption{Confidence contours for X7 obtained using the {\tt steppar} tool (discussed in Appendix \ref{subsection:steppar}). }
\end{figure}

\begin{figure}[ht]
\centering
    \includegraphics[scale=0.6]{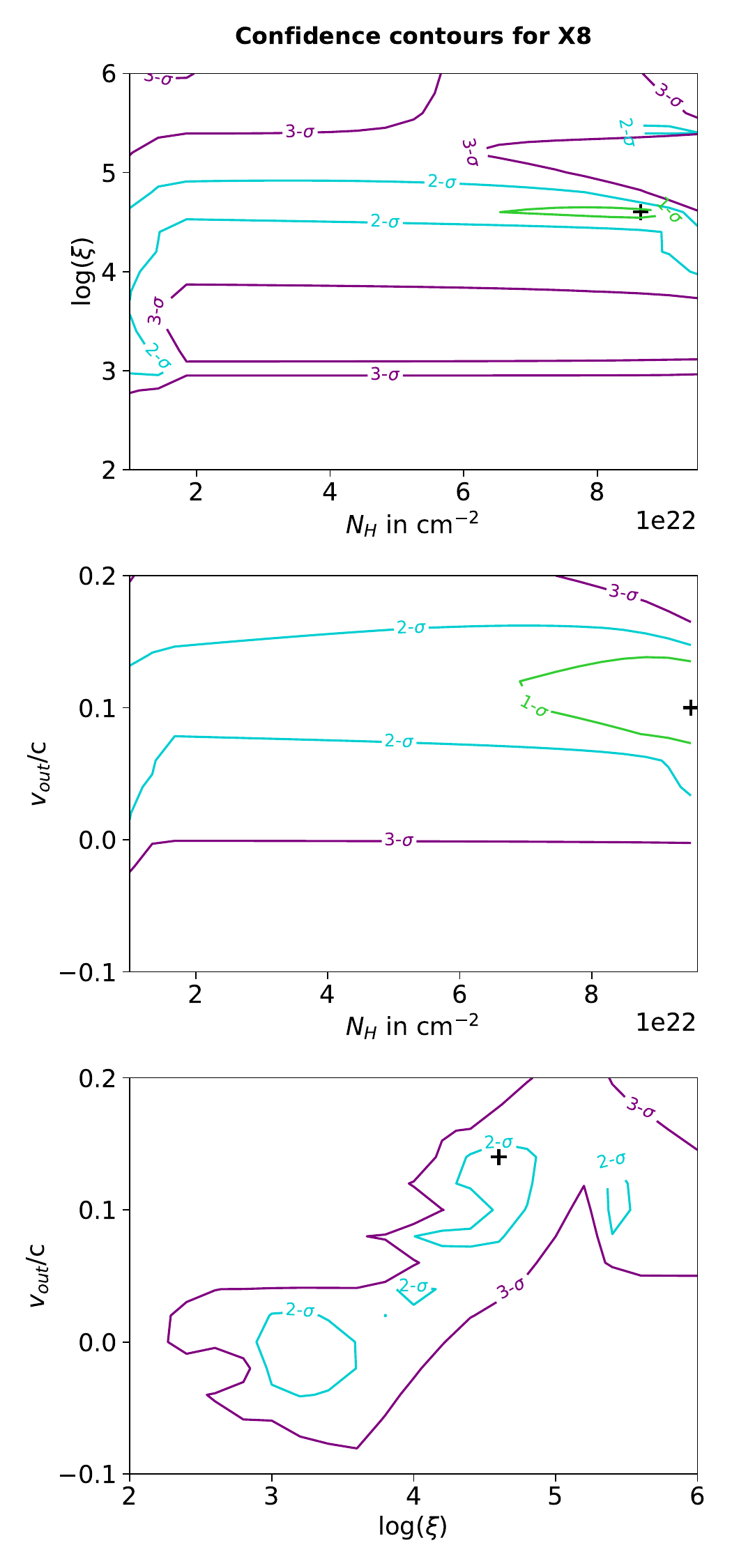} 
    \caption{Confidence contours for X8 obtained using the {\tt steppar} tool (discussed in Appendix \ref{subsection:steppar}). Note: the 1-$\sigma$ contour for $log(\xi)$ vs $v_{out}/c$ (bottom panel) is smaller than the size of the best-fit parameters idicated using the ``+" marker and is not visible here. }
\end{figure}

\clearpage
\section{Supplementary tables}
\begin{table}[ht]
\centering
\begin{tabular}{|c||c|c|c|c|} 
 \hline
   & Temperature (keV) & Luminosity (ergs s$^{-1}$) & nH ($10^{22}$ cm$^{-2}$)& C-stat/d.o.f\\
 \hline\hline
 X1 & $0.059\pm0.001$ & $44.70\pm0.01$ & $0.017\pm0.004$ & 39.3/17\\ 
 \hline
 X2 & $0.056\pm0.001$ & $44.81\pm0.08$ & $0.028\pm0.006$ & 36.4/17\\
 \hline
 X3 & $0.057\pm0.001$ & $44.83\pm0.06$ & $0.032\pm0.005$ & 52.5/17\\
 \hline
 X4 & $0.048\pm0.001$ & $44.58\pm0.11$ & $0.028\pm0.006$ & 30.0/17\\
 \hline 
 X5 & $0.045\pm0.001$ & $44.60\pm0.09$ & $0.032\pm0.005$ & 33.1/17\\ 
 \hline
 X6 & $0.044\pm0.002$ & $44.58\pm0.20$ & $0.029\pm0.010$ & 15.3/16\\ 
 \hline
 X7 & $0.037\pm0.002$ & $44.82\pm0.25$ & $0.050\pm0.012$ & 44.9/13\\ 
 \hline
 X8 & $0.036\pm0.004$ & $44.75\pm0.55$ & $0.053\pm0.029$ & 24.5/13\\ 
 \hline
 X9 & $0.039\pm0.001$ & $43.81\pm0.09$ & $0^{*}$ & 13.4/14\\ 
 \hline
 X10 & $0.037\pm0.003$ & $43.71\pm0.23$ & $0^{*}$ & 4.6/6\\  
 \hline 
 X11 & $0.039\pm0.003$ & $43.23\pm0.19$ & $0^{*}$ & 10.8/12\\ 
 \hline
 X12 & $0.038\pm0.004$ & $43.23\pm0.23$ & $0^{*}$ & 7.8/12\\ 
 \hline
 X13 & $0.035\pm0.002$ & $43.31\pm0.14$ & $0^{*}$ & 16.5/16\\  [1ex]
 \hline
\end{tabular}
\caption{Best-fit parameter values of temperature, model predicted integrated luminosity (13.6 eV to 13.6 keV), and host neutral column (nH) for the disk blackbody model. The ``*'' indicates frozen value of {zTBabs} neutral column in X9-13 since the best-fit nH, when left free, was consistent with 0 cm$^{-2}$.}
\label{table:diskbb}
\end{table}

\begin{table}[ht]
\centering
\begin{tabular}{|c||c|c|c|c|} 
 \hline
   & Temperature (keV) & Luminosity (ergs s$^{-1}$) & nH ($10^{22}$ cm$^{-2}$)& C-stat/d.o.f\\
 \hline\hline
 X1 & $0.065\pm0.002$ & $44.54\pm0.08$ & $0.008\pm0.005$ & 10.3/14\\ 
 \hline
 X2 & $0.061\pm0.002$ & $44.64\pm0.11$ & $0.017\pm0.008$ & 16.1/14\\
 \hline
 X3 & $0.063\pm0.002$ & $44.59\pm0.10$ & $0.014\pm0.008$ & 6.4/14\\
 \hline
 X4 & $0.048\pm0.002$ & $44.57\pm0.11$ & $0.028\pm0.006$ & 29.3/16\\
 \hline 
 X5 & $0.049\pm0.004$ & $44.31\pm0.18$ & $0.019\pm0.009$ & 20.0/14\\ 
 \hline
 X6 & $0.045\pm0.003$ & $44.61\pm0.21$ & $0.031\pm0.011$ & 13.2/15\\ 
 \hline
 X7 & $0.054\pm0.006$ & $44.12\pm0.53$ & $0^{*}$ & 11.0/11\\ 
 \hline
 X8 & $0.056\pm0.009$ & $43.69\pm0.22$ & $0^{*}$ & 14.2/11\\
 \hline
\end{tabular}
\caption{Best-fit parameter values of temperature, model predicted integrated luminosity (13.6 eV to 13.6 keV), and host neutral column (nH) for {disk blackbody + UFO (XSTAR) model}. The ``*'' indicates frozen value of {zTBabs} neutral column in X7 and X8 since nH was consistent with 0 cm$^{-2}$.}
\label{table:ufo_diskbb}
\end{table}

\begin{table}[ht]
\centering
\begin{tabular} {|c||c|c|c|}
    \hline
        & Column density in atoms cm\textsuperscript{\textminus 2} & log($\xi$) in log(erg cm s\textsuperscript{\textminus 1}) & $v_{\rm out}/c$\\
    \hline\hline
     X1 & $1.58^{+0.72}_{-0.55} \times 10^{21}$ & $2.36_{-0.30}^{+0.68}$ & $-0.19_{-0.02}^{+0.02}$\\
     \hline
     X2 & $3.76_{-2.29}^{+1.78} \times 10^{21}$ & $< 3.78$ & $-0.17_{-0.06} ^{+0.05}$\\
     \hline
     X3 & $2.15_{-0.57}^{+3.21} \times 10^{21}$ & $2.81_{-0.35}^{+0.55}$ & $-0.20_{-0.02}^{+0.03}$ \\
     \hline
     X4 & $<7.72 \times 10^{20}$ & $2.00^{*}$ & $0.00^{*}$\\
     \hline 
    X5 & $3.29_{-1.95}^{+3.79} \times 10^{21}$ & $4.81_{-2.18}^{+1.18}$ & $< -0.15$ \\
     \hline
    X6 & $<2.06 \times 10^{21}$  & $2.00^{*}$ & $0.00^{*}$ \\
    \hline
    X7 & $5.56_{-4.27}^{+3.21} \times 10^{22}$ & $4.97_{-0.37}^{+0.37}$ & $< -0.02$\\ 
    \hline
    X8 & $1.25_{-0.63}^{+0.28} \times 10^{22}$ & $3.06_{-0.14}^{+1.49}$ & $< -0.05$\\
    \hline
    \end{tabular}
\caption{Best-fit XSTAR parameters with {disk blackbody + UFO (XSTAR) model} for X1-8. Parameter values for log($\xi$) and $v_{\rm out}/c$ for X4 and X6 were frozen to their hard limits (indicated using ``*'') so as to obtain upper limits on their respective $N_{H}$ ionized columns.}
\label{table:5a}
\end{table}

\clearpage
\section{Supplementary figures}
\begin{figure}[ht]
    \includegraphics[scale=0.5]{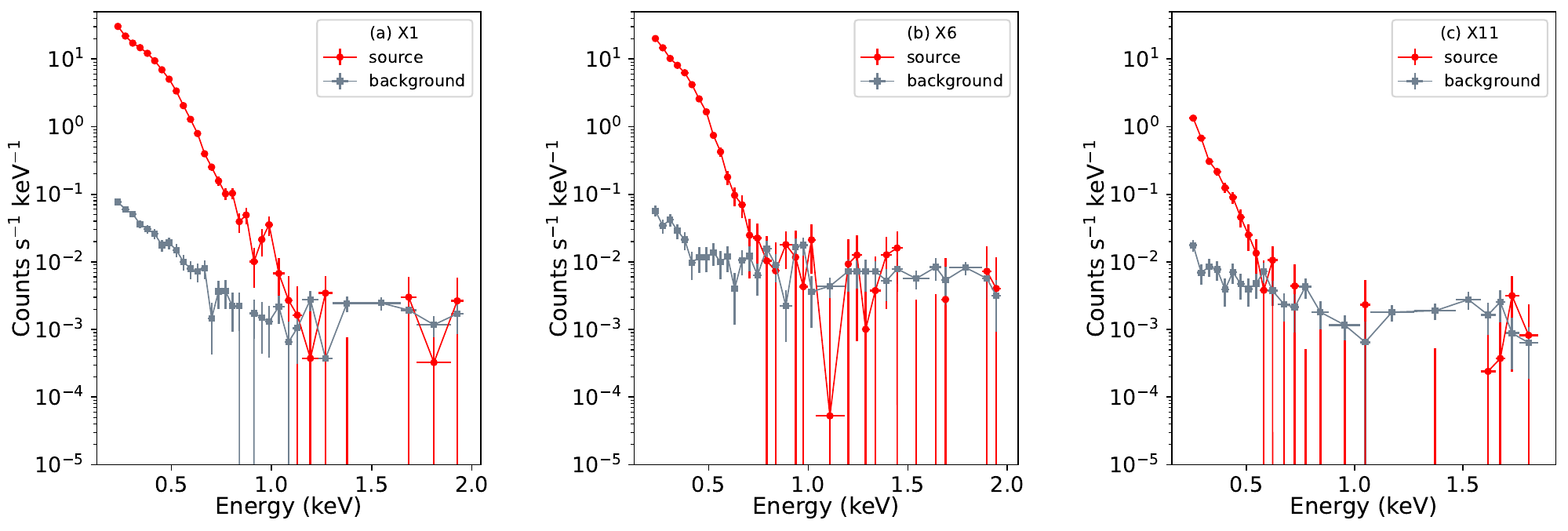}
    \caption{Source and background spectra in the energy range 0.2-2 keV for (a) X1, where the background spectrum dominates beyond 1.2 keV; (b) X6, where the background spectrum dominates beyond 1 keV; and (c) X11, where the background spectrum dominates beyond 0.6 keV}
\label{fig:srcbkg}
\end{figure} 
\begin{figure}[ht]
    \centering
    \includegraphics[width=0.55\textwidth]{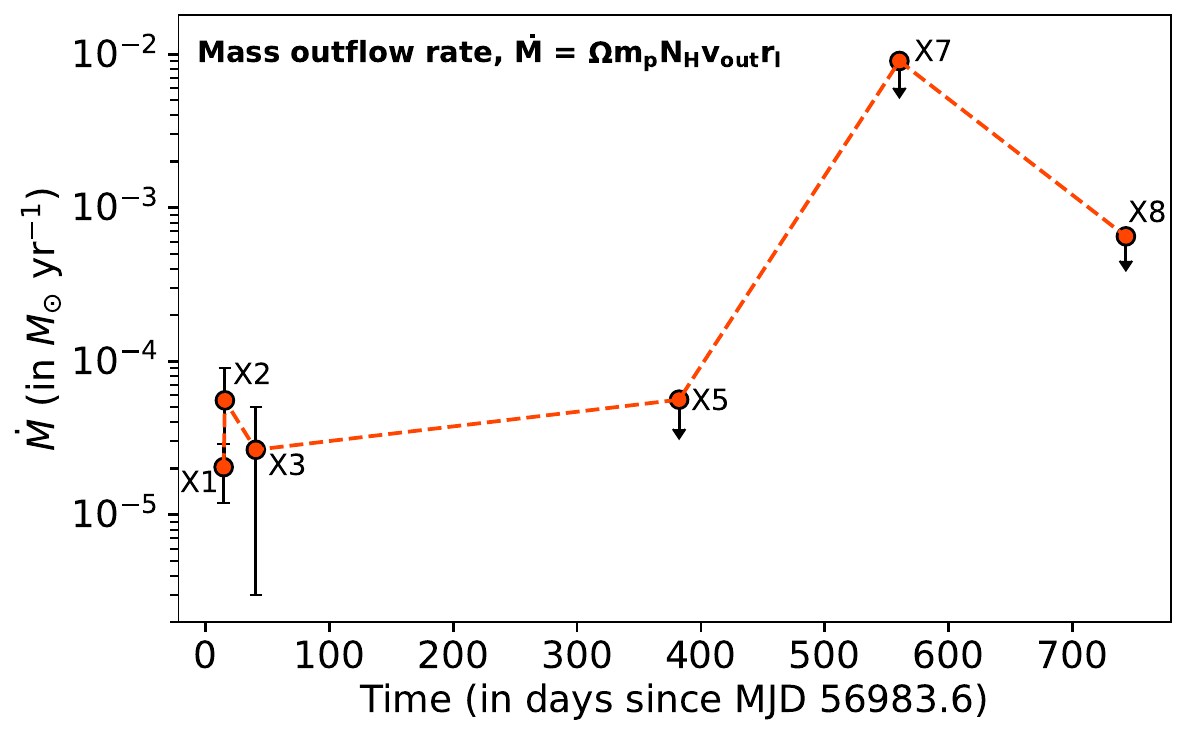} \\  
    \caption{Mass outflow rate in observations where outflows were detected, computed using $\dot{M} = \Omega m_{p} N_{H} v_{out} r_{l}$ (see main text for details). The mass outflow rate increases in the late-time observations X7-8 by over one magnitude. Errors for $N_{H}$, $v_{out}$, and $M_{BH}$ have been included (with log($M_{BH}$/\(\textup{$M$}_\odot\))$=6.23\pm0.40$). The $\dot{M}$ upper limits for X5, X7, and X8 shown with downward arrows were computed considering an upper limit on their respective velocities.}
    \label{fig:mass_rates}
\end{figure}

\end{document}